\documentclass[a4paper,11pt,headings=big,DIV=12]{scrartcl}
\pdfoutput=1
\usepackage{amsmath,amssymb}
\usepackage{color,xcolor}
\definecolor{blue}{rgb}{0,0,0.5} 
\usepackage[colorlinks,linkcolor=blue,urlcolor=blue,citecolor=blue]{hyperref}
\usepackage{graphicx}
\addtokomafont{disposition}{\rmfamily\boldmath}
\usepackage[utf8]{inputenc}
\usepackage{enumerate} 
\usepackage{slashed}
\usepackage{cite} 
\usepackage{comment}
\usepackage{multirow}
 \usepackage{bbold}
\allowdisplaybreaks
\newcommand{\eom}{\text{eom}}

\newcommand{\ba}{\be_a}
\newcommand{\bb}{\be_b}
\newcommand{\bc}{\be_c}
\newcommand{\barb}{\bar{b}}
\newcommand{\barbb}{\bar{\bar{b}}}
\newcommand{\GM}{{\cal G}}
\newcommand{\AS}{ \tilde{\cal G}}
\newcommand{\ASz}{\tilde{\chi}}

\newcommand{\ONE}{\mathbb{1}}

\newcommand{\zam}[2]{\chi_{#1 #2}}
\newcommand{\zamS}[2]{\chi^{\MOM}_{#1 #2}}

\newcommand{\zamP}[2]{\chi'_{#1 #2}}
\newcommand{\zamMS}[2]{\chi^{\MS}_{#1 #2}}

\newcommand{\zamR}[3]{\chi_{#1 #2}^{#3}}

\newcommand{\Ra}{  { \RR_1} }
\newcommand{\RR}{  { {\cal R}} }
\newcommand{\Rb}{  { \RR_2} }
\newcommand{\RB}{  { \RR_{b}} }
\newcommand{\RT}{  { \RR_{\chi}} }

\newcommand{\rnDI}[3]{r_{#1 #2 }^{\ONE (#3)}}

\newcommand{\CW}{{\mathbb C}}
\newcommand{\fin}{[\text{finite}]}

\newcommand{\MS}{{ \textrm{MS}}}
\newcommand{\Rs}{  {R^2} }
\newcommand{\Nscheme}{\text{MOM}}
\newcommand{\NScheme}{\text{MOM}}
\newcommand{\Rsscheme}{$R^2$}
\newcommand{\rsscheme}{R^2}

\newcommand{\pas}{ \partial_{ \als}}
\newcommand{\als}{{a_s}}

\newcommand{\alsIR}{{a_s^\IR}}
\newcommand{\alsp}[1]{{a_s^{\;#1}}}

\newcommand{\Ntt}{\ONE}
\newcommand{\oldC}{\CW}

\newcommand{\Mo}{M}

\newcommand{\momT}[1]{{\Mo}_{#1}}
\newcommand{\momDD}[1]{\Mo_{#1#1}}

\newcommand{\cwi}[2]{\oldC_{#1 #2}^{\Ntt}}
\newcommand{\cwiR}[3]{\oldC_{#1 #2}^{\Ntt, #3 }}
\newcommand{\cwiDR}[2]{\oldC_{#1 #1}^{\Ntt, #2}}

\newcommand{\cwiD}[1]{\oldC_{#1 #1 }^{\Ntt}}
\newcommand{\cwiDS}[1]{\oldC_{#1 #1 }^{\Ntt,\MOM}}

\newcommand{\MOM}{{\textrm{MOM}}}

\newcommand{\lD}[1]{L_{#1 #1}^{\Ntt,\RR}}
\newcommand{\lDS}[1]{L_{#1 #1}^{\Ntt,\MOM}}
\newcommand{\lDMS}[1]{L_{#1 #1}^{\Ntt,\MS}}
\newcommand{\LbMS}{L_b^{\MS}}

\newcommand{\bebMS}{\be_b^{\MS}}

\newcommand{\lnD}[2]{L_{#1 #2}^{\Ntt,\RR}}

\newcommand{\lDR}[2]{ L_{#1 #1}^{\Ntt,#2}}

\newcommand{\lnDR}[3]{ L_{#1 #2}^{\Ntt,#3}}

\newcommand{\ZopI}[2]{{ \Zop_{#1}^{\phantom{#1} #2}}  }

\newcommand{\Zop}{{\mathbb{Z}}}

\newcommand{\gm}{\mathrm{g}}

\newcommand{\TEMT}[1]{ T^{#1}_{ \;\;#1}}
\newcommand{\TEMTO}{\Theta}

\newcommand{\vev}[1]{\langle #1 \rangle}

\newcommand{\matel}[3]{\langle #1|#2|#3\rangle}
\newcommand{\al}{\alpha}
\newcommand{\be}{\beta}
\newcommand{\ga}{\gamma}
\newcommand{\de}{\delta}
\newcommand{\la}{\lambda}
\newcommand{\eps}{\epsilon}

\newcommand{\Rtr}[2]{\Theta}

\newcommand{\UV}{{\textrm{UV}}}
\newcommand{\IR}{{\textrm{IR}}}

\newcommand{\Zpart}{{\cal Z}}

\newcommand{\kCBZ}{\kappa}

\newcommand*{\mathcolor}{}
\def\mathcolor#1#{\mathcoloraux{#1}}
\newcommand*{\mathcoloraux}[3]{%
  \protect\leavevmode
  \begingroup
    \color#1{#2}#3%
  \endgroup
}

\begin{document}

\begin{flushright}
\begin{tabular}{l}
CP3-Origins-2016-057 DNRF90 \\
 \end{tabular}
\end{flushright}
\vskip1.5cm

\begin{center}
{\Large\bfseries \boldmath On the  Flow of the $\Box R$ Weyl-Anomaly}\\[0.8 cm]
{\Large%
Vladimir Prochazka
and Roman Zwicky,
\\[0.5 cm]
\small
 Higgs Centre for Theoretical Physics, School of Physics and Astronomy,\\
University of Edinburgh, Edinburgh EH9 3JZ, Scotland 
} \\[0.5 cm]
\small
E-Mail:
\texttt{\href{mailto:v.prochazka@ed.ac.uk}{v.prochazka@ed.ac.uk}},
\texttt{\href{mailto:roman.zwicky@ed.ac.uk}{roman.zwicky@ed.ac.uk}}.
\end{center}
  
\bigskip
\pagestyle{empty}

\begin{abstract}

An important aspect of Weyl anomalies is that 
they encode information on the irreversibility of the renormalisation 
group  flow.  
We consider, $\Delta \bar b = \bar{b}^{\textrm{UV}} - \bar{b}^{\textrm{IR}}$, the difference of
the ultraviolet and  infrared value of the $\Box R$-term 
of the Weyl anomaly.  The quantity is related to the fourth moment of the trace of the
energy momentum tensor correlator  for theories  which 
are conformal at both ends. Subtleties arise for non-conformal fixed points
 as might be the case for  infrared fixed points with broken chiral symmetry.
Provided that the moment converges, $\Delta \bar{b}$ is then automatically positive by unitarity. 
Written as an integral over the renormalisation scale, flow-independence follows 
since its integrand is a total derivative. 
Furthermore, using a momentum subtraction scheme (MOM) the 4D 
Zamolodchikov-metric is shown to be strictly positive beyond perturbation theory  and equivalent to the metric of a conformal 
manifold at both ends of the flow. 
In this scheme $\bar{b}(\mu)$ can be extended outside
the fixed point to a monotonically decreasing function. 
The ultraviolet finiteness of the fourth moment enables us to define a   scheme for the $\de {\cal L}  \sim    b_0 R^2$-term, for which the $R^2$-anomaly vanishes along the flow. 
In the  MOM- and the $R^2$-scheme, 
 $  \bar{b}(\mu) $ 
 is shown to satisfy a gradient flow type equation.  
We verify our findings in free field theories, higher derivative theories and extend $\Delta \barb$  
and the Euler flow $\Delta \be_a$ 
for a Caswell-Banks-Zaks fixed point for QCD-like theories to next-to-next-to leading order using a recent $\vev{G^2G^2}$-correlator computation.

 \end{abstract}

\newpage

\setcounter{tocdepth}{3}
\setcounter{page}{1}
\tableofcontents
\pagestyle{plain}

\newpage

\section{Introduction}
\label{sec:intro}
 
It is well-known that moments of the correlator of the trace of  the energy 
momentum tensor (TEMT)   provide information on the flow of  Weyl anomalies in theories with an ultraviolet (UV) and an infrared (IR) conformal fixed point (FP).
For example the 2D Weyl anomaly $\vev{\TEMT{\rho}}_{\textrm{CFT}} = - (\be_c/(2 4 \pi)) R$,  
where $\TEMT{\rho}$ is the TEMT, 
$R$ the Ricci scalar and $\be_c =1$ for a free scalar field, is probed by the second moment
\begin{equation}
\label{eq:mom2D}
\Delta \be_c^{2D} = \be_c^\UV -  \be_c^\IR = 3 \pi \int d^2 x \, x^2  \vev{ \TEMTO(x) \TEMTO(0)}_c 
\geq 0 \;.
\end{equation}
of the TEMT in flat space $\TEMT{\rho}|_{\textrm{flat}} \to \TEMTO$.
This formula is  Cardy's version \cite{Cardy:1988tj} of 
the celebrated $c$-theorem \cite{Zamolodchikov:1986gt} and  
$\vev{\dots}_c$ stands for the connected component of the vacuum expectation value (VEV).
Positivity follows  reflection positivity  \cite{Cardy:1988tj} 
or  the positivity of the spectral representation \cite{Cappelli1}. 
In 3D there are no Weyl anomalies on dimensional grounds but a relation 
analogous to \eqref{eq:mom2D} exists for the moment of two electromagnetic currents 
 related to the flow of the parity anomaly \cite{anomaly1}.

In this work we exploit the finiteness conditions for $2$-functions, worked 
out in a previous paper \cite{PZ16}, to obtain results on 4D Weyl-anomalies.  
In 4D an analogous relation to  \eqref{eq:mom2D} has been proposed in 
\cite{Cappelli1,Cappelli2,A99} and 
indirectly in \cite{Zee81},
 \begin{equation}
\label{eq:mom4D}
\Delta \barb =  \barb^\UV -  \barb^\IR =  \frac{1}{2^9\, 3} 
\int d^4 x\, x^4   \vev{ \TEMTO(x) \TEMTO(0)}_c  \geq 0 \;.
\end{equation}
where $\barb $ is the $\Box R$-term \cite{A99} of the Weyl or conformal anomaly 
\cite{birrell1982quantum}\footnote{In this paper the coefficients  in front of the 
geometric invariants are denoted by $\be$-functions, 
in (dis)accordance with  \cite{JO90, H81,S16} (\!\!\cite{KS11,Prochazka:2015edz}).
The association of the letters $a,b$ and $c$  
with the geometric invariants is variable in the literature. 
Our notation follows Shore's review \cite{S16} in this respect.}$^{,}$\footnote{The  cosmological constant $\Lambda^\IR$  may or may not be tuned  to  zero by an appropriate 
UV-counterterm. Note that the parametrisation 
$\vev{T_{\al \be}}=  g_{\al \be} \Lambda^\IR + \dots $ is being used.
In QCD-like theories, for example, the cosmological constant receives contributions from the 
gluon condensate  
$\Lambda^\IR(\mu) = \be(\mu)/2 \vev{G^2}_{\mu}$-term. This term is essential for    
$ \frac{d}{d \ln \mu} \vev{\TEMT{\rho} } = 0$ cf. section \ref{sec:R2QCD}.}  
\begin{equation}
\label{eq:VEVTEMT}
\vev{\TEMT{\rho}(x)} =  \frac{1}{\sqrt{\gm}} \left(- {\de_{s(x)} } \right) \ln \Zpart   =   -( \ba^\IR E_4 + \bb^\IR H^2 + \bc^\IR W^2  )  + 4 \barb^\IR \Box  H 
+ 4 \Lambda^\IR  \;,
\end{equation}
and $H$ is the commonly used shorthand \cite{H81,JO90} 
$$H \equiv \frac{1}{(d-1)} R \;.$$
Above $\de_{s(x)} \equiv \frac{\de}{\de s (x)}$ under  $ \gm_{\al \be} \to e^{-2 s(x) } \gm_{\al\be}$,  
and  $E_4$, $W^2$ and $R$ are the Euler, the Weyl squared  and the Ricci scalar. 
 The superscript IR indicates that all modes have been integrated out. 
The quantities 
 $\be_{a,b,c}^\IR$ are scheme-independent and   determined by the IR-theory. 
 In the case where the IR-theory is a CFT, $\be_{b}^{\textrm{CFT}} = 0$  \cite{D77,BCR83} 
 implies that  $\be_{a,c}^{\textrm{CFT}} \neq 0$ are to be regarded 
 as the true Weyl  anomalies. 
 Turning our attention to the $\barb^\IR$-term, the first thing to notice is
that  this term  shifts linearly  when adding local term to the UV-action 
(conventions as in  \cite{PZ16})
 \begin{equation}
 \label{eq:amb}
  {\cal L}^\UV  \to   {\cal L}^\UV +  \frac{1}{8} \omega_0 H^2  \;, \quad  \barb^\IR \to \barb^\IR - \frac{1}{8}\omega_0 \;.
 \end{equation}
This is presumably related to regularisation dependence found in explicit computations (e.g. \cite{birrell1982quantum,D77,BC77,AGS05,VFGS15,Chu16}).
   The $\Box R$-term is therefore sometimes viewed as not being meaningful. 
  One of the main points of our paper is that in physical meaningful quantities, such as 
  $\Delta \barb$ \eqref{eq:mom4D},  this ambiguity has to cancel. 
  For the flow,  $\omega_0$ is merely 
  to be seen as the initial condition which does not affect the difference $\Delta \barb$.
   A valuable result of this paper is that we show that  
   the $\Box R$ flow is given by, 
  \begin{equation}
  \label{eq:barbdmu}
  \Delta \barb = \frac{1}{8} \int_{-\infty}^{\infty} \zamR{\theta}{\theta}{\RT}(\mu')  d \ln \mu'  \;, 
    \end{equation}
 an integral over $\zamR{\theta}{\theta}{\RT} $, 
 the scale derivative of the renormalised  $\vev{\TEMTO\TEMTO}^\RT$-correlator. 
  In particular we identify a \Nscheme-type scheme for which  
   \begin{equation}
   \label{eq:barbdmu2}
  \zamS{\theta}{\theta}(\mu) =  \zamS{A}{B}(\mu)  \be^A(\mu) \be^B(\mu) \;,  \qquad   \zamS{A}{B}(\mu) \geq 0 \;,
  \end{equation}
  with $\zamS{A}{B}(\mu)$, the 4D analogue of
  the Zamolodchikov-metric,  being a positive definite 
  matrix  along the flow.   
Since $ \zamS{\theta}{\theta}$ can be written as a $\ln \mu$-derivative 
it follows that $\Delta \barb$ is flow-independent. The positivity of $ \zamS{A}{B}$ allows 
us to define a $ \barb (\mu)^\MOM$ outside the FPs as a monotonically decreasing function.  
Building on the observation that $\Delta \barb $ is UV-finite for asymptotically-free and 
asymptotically-safe flows \cite{PZ16}, $\Delta \barb \geq 0$ follows from the spectral representation.  
Furthermore, finiteness  
 allows us to define a scheme ($R^2$-scheme) for which the $R^2$-anomaly vanishes along the flow. In this  scheme  $\barbb \equiv  \barb (\mu)^\MOM_{R^2}$ is shown to 
obey a gradient flow  type equation.  
  Furthermore we provide  $\Delta \bar b$  to NNLO in QCD-like theories  using a recent NNLO 
  computation of the $\vev{G^2 G^2}$-correlator. The latter  also extendeds 
  to the Euler flow  $\Delta \be_a$ since it is proportional to $\Delta \barb$ up to NNLO 
  around   the Caswell-Banks-Zaks (CBZ) FP.


The paper is organised as follows.  The flow properties of $\Box R$ are presented in section \ref{sec:boxR} with the \Nscheme- and \Rsscheme-scheme  for the $\vev{\TEMTO \TEMTO }$-correlator and the  $b$-coupling defined in sections \ref{sec:Gscheme} and \ref{sec:R2} respectively.  The main part consists of the description of the properties of 
$\Delta \barb$,  $\barb^\RR(\mu)$  and the Zamolodchikov-metric $\zam{A}{B}^\RR$ in section   \ref{sec:prop} followed by a discussion of the IR- and UV-convergence of the correlator indicating potential limitations.
 Section \ref{sec:boxR}, which can  be considered as the main part of the paper, is summarised in section \ref{sec:summary}.
  The explicit scheme change of the Zamolodchikov-metric from 
  the \Nscheme- to the $\MS$-scheme, $\Delta \barb$ at a CBZ-FP and
  the renormalisation of $G^2$ in the \Rsscheme-scheme are discussed in sections
  \ref{sec:schemeC}, \ref{sec:BZ} and \ref{sec:R2QCD} respectively. 
  Free  field theory computation of scalars and fermions are presented in section \ref{sec:free} and 
  a free higher derivative computation is deferred to appendix \ref{app:higher}. 
  Three derivations 
  of \eqref{eq:mom4D} using anomalous Ward identities (WI), an anomaly matching argument and an explicit derivation 
  in QCD-like theories are provided in appendices  \ref{app:AWI}, \ref{app:athm} and  \ref{app:QCD-like} respectively. The antisymmetric part of the gradient flow equation is elaborated on in appendix \ref{app:asym}.
  Comments on different ways of handling the gravity counterterms are discussed in appendix  
   \ref{app:2-ways} followed by the computation of $\Box R$-flow in a higher derivative theory in 
   appendix \ref{app:higher}.
  Conventions of the QCD $\be$-function and the CBZ-FP are specified in appendix 
  \ref{app:beta}.

\section{The Flow of $\Box R$ (or $\Delta \barb$)}
\label{sec:boxR}

Before writing  the fourth moment \eqref{eq:mom4D} in terms of the 
4D Zamolodchikov-metric and showing  positivity, monotonicity and the gradient flow type property,  
we specify some  definitions, notations and assumptions of this paper.  
We work with the conventions of a Euclidean field theory  and assume the operator-part of the TEMT 
to be of the form  $\TEMTO = \be^Q [O_Q]$ (summation over $Q$ is implied).   
We refer the reader our previous work \cite{PZ16} regarding the terminology of the TEMT  
$\TEMT{\rho}$
which splits into an operator, equation of motion and gravity part. 
At the exception of the free field theory examples in section \ref{sec:free} dimensionless  couplings 
are assumed. The bare interaction Lagrangian is parameterised by  
${\cal L} =  g_0^Q O_Q$, $g^Q_0$ are couplings and $[O_Q] = \ZopI{Q}{P}  O_P$ denote  renormalised (composite) 
operators defined by the  local quantum action principle (QAP)  
$\vev{[O_A(x)]} =  \left( -  \de_{A(x)} \right) \ln \Zpart$ where $\Zpart$ and  $\de_{A(x)} \equiv \frac{\de}{\de g^A(x)}$ are 
the partition function and the variational derivative of the localised coupling respectively. 
Curved space is a tool  to expose the Weyl anomalies \eqref{eq:VEVTEMT} at  
lower-point functions  and has no further physical meaning in this work.

The object of study is the $\vev{\TEMTO\TEMTO}$-correlator ($\TEMTO = [\TEMTO] $  since it is physical e.g. \cite{Adler:1976zt,Nielsen:1977sy})\,\footnote{The restrictive 
structure of \eqref{eq:Gamma(p2)} follows from the flat-space translational WI 
$\int d^4 x e^{i p \cdot x} \vev{\TEMTO_{\al \be} (x)\TEMTO_{\ga \de}(0)}_c = P^{(0)}_{\al \be \ga \de} \Gamma^{(0)} + 
P^{(2)}_{\al \be \ga \de}   \Gamma^{(2)} + P^{(CT)}_{\al \be \ga \de}  \vev{ \TEMTO}$.  
From  the traces of the spin $0$ and $2$ structures,  $P^{(0)}_{\al \ga \al   \ga  } \sim p^4$  and $P^{(2)}_{\al \ga \al   \ga  }=0$ (note $P^{(CT)}_{\al \ga \al   \ga  } = c_T$),  one infers that 
   $ \Gamma^{(0)}(p) \sim \cwiD{\theta} (p)  $.}  
\begin{equation}
\label{eq:Gamma(p2)}
\Gamma_{\theta \theta}(p) =  \int d^4x \, e^{i x \cdot p}   \vev{ \TEMTO(x) \TEMTO(0)}_c 
=  \cwiD{\theta} (p)  p^4  +   c_T  \vev{ \TEMTO}  
\end{equation}
for which $\cwiD{\theta} (p)$ is UV-finite \cite{PZ16} and IR-finite for $p >0$  with subtleties 
for $p \to 0$ for theories spontaneously broken chiral symmetry  to be discussed in section \ref{sec:chiral}.
The contact term (CT) $c_T$  is of no relevance for the flow itself, the scalar product is defined 
as usual $a \cdot b \equiv  a_\al b^\al$ and $p \equiv \sqrt{ p \cdot p}$.
Defining  
\begin{eqnarray}
\label{eq:4mom}
\momDD{\theta}^{(2)}(p)  =  \hat{P}_2  
    \Gamma_{\theta \theta}(p)  =
 \frac{1}{2^6 3}  \int d^4x \, e^{i x \cdot p}  x^4  \vev{ \TEMTO(x) \TEMTO(0)}_c \;,
\end{eqnarray}
with 
\begin{equation}
\label{eq:P2}
\hat{P}_2 =\frac{1}{2^6 3} ( \partial_{p_\al} \partial_{p^\al})^2 \;, \quad \hat{P}_2 \, p^4 =1 \;,
\end{equation}
the fourth moment \eqref{eq:mom4D} is then proportional to  $\momDD{\theta}^{(2)}(0)$. The bare quantities 
$\momDD{\theta}^{(2)}(p)$ and $ \cwiD{\theta} (p) $ satisfy unsubtracted 
K\"all\'en-Lehmann spectral/dispersion representations  of the form\,\footnote{\label{foot:disp}
The dispersion  relation for the correlation function \eqref{eq:Gamma(p2)}
reads $\Gamma^\RR_{\theta\theta}(p)  = \int_0^\infty ds  \frac{\rho(s)}{s + p^2}    +   \omega^\RR_4(\mu) + 
\omega_2^\RR(\mu) p^2 +  \omega_0 p^4$.  The constants $\omega^\RR_{2,4}(\mu)$ take care
of the quadratic and quartic divergences whereas the logarithmic part is convergent and $\omega_0$ 
is therefore a true constant independent of $\mu$. 
Eq.~\eqref{eq:Cdisp} and \eqref{eq:Mdisp} are  obtained from the ones above by using  $p^4 \cwiD{\theta}(p )  = \Gamma_{\theta\theta}(p) - 
\Gamma_{\theta\theta}(0) - p^2 \frac{d}{d p^2} \Gamma_{\theta\theta}(0)$ and $\momDD{\theta}^{(2)}(p)  =  \hat{P}_2  
    \Gamma_{\theta \theta}(p) $.}
\begin{eqnarray}
\label{eq:Cdisp}
\cwiD{\theta}(p)  &\; =\; & \int_0^\infty \frac{d s}{s^2}  \frac{\rho(s)} {(s + p^2)}  + \cwiD{\theta}(\infty) \;, \\[0.1cm]
\label{eq:Mdisp}
\momDD{\theta}^{(2)}(p) &\; =\; &
 \int_0^\infty d s  \frac{  \, s(s-p^2)\rho(s) }{(s + p^2)^5}  + \momDD{\theta}^{(2)}(\infty) \;,
\end{eqnarray}
  where the spectral function $\rho(s)$  is of mass dimension four 
and defined as a formal sum over  a complete set of spin $0$ physical states,
\begin{equation}
\label{eq:spectral}
\rho(s)  = (2 \pi)^3 \sum_n \theta((p_n)_0) \de(p_n^2 -  s) | \matel{n(p_n)}{\TEMTO}{0}|^2   \geq 0 \;,
\end{equation}
 with $p_n$ denoting 
momenta in Minkowski-space 
and  $ \theta(x)$  is the step-function. 

From the representations eqs.~(\ref{eq:Cdisp},\ref{eq:Mdisp}) one deduces that 
$\momDD{\theta}^{(2)}(p) - \cwiD{\theta}(p) $ is a finite $p$-dependent function 
for which $ \momDD{\theta}^{(2)}(0) - \momDD{\theta}^{(2)}(\infty)    = \cwiD{\theta}(0) -\cwiD{\theta}(\infty) $
holds. Furthermore, using 
 and  with 
eqs.(\ref{eq:Gamma(p2)},\ref{eq:4mom}) it follows
that 
\begin{equation}
\label{eq:CM}
\momDD{\theta}^{(2)}(0) = \cwiD{\theta} (0) \;, \quad \momDD{\theta}^{(2)}(\infty) = \cwiD{\theta} (\infty) \;.
\end{equation}
Together with  \eqref{eq:mom4D} this implies
\begin{equation}
\label{eq:state}
\Delta \barb =    \frac{1}{8} \momDD{\theta}^{(2)}(0)  \;,
\end{equation}
which modifies to   $\Delta \barb = \frac{1}{8}( \momDD{\theta}^{(2)}  (0)  - \momDD{\theta}^{(2)} (\infty))$  in the case where there is a finite contribution at 
infinity. This is  for instance necessary  when adding a term \eqref{eq:amb} to the UV Lagrangian as discussed 
towards the end of section \ref{sec:scheme}. 
 The reason for introducing $\cwiD{\theta}(p)$ is 
 that, contrary to $\momDD{\theta}^{(2)}(p)$,  it is monotonic in $p$ 
  allowing us to define a 
positive Zamolodchikov-metric in the $\MOM$-scheme (cf. section \ref{sec:scheme}).
We stress that 
$ \momDD{\theta}^{(2)}(0)$ is a bare,  $\mu$-independent, quantity and in the case where it
is IR- and UV-finite (cf. section \ref{sec:convergence})  it therefore qualifies as a physical 
observable. Three different derivations of \eqref{eq:mom4D} are given in appendices 
\ref{app:AWI}, \ref{app:athm} and  \ref{app:QCD-like}.

\subsection{Generic Scheme-definition  for the $\vev{\TEMTO\TEMTO}$-Correlator}
\label{sec:Gscheme}

In order to perform an RG-analysis,   
 the bare term in \eqref{eq:4mom}  
 is written as a sum of a renormalised term and a counterterm, 
\begin{equation}
\label{eq:scheme1}
 \momDD{\theta}^{(2)} (g^Q(p/\mu_0)) = \momDD{\theta}^{(2),\RR}  (p/\mu,g^Q(\mu/\mu_0))+  
  \lD{\theta}(g^Q(\mu/\mu_0)) \;.
  \end{equation} 
Above $\mu_0$ is some reference scale and 
$ \momDD{\theta}^{(2),\RR}(p/\mu, \als(\mu)) = \momDD{\theta}^{(2),\RR}(p/\mu_0, \mu/\mu_0, \als(\mu))$
but  most of the time   $\momDD{\theta}^{(2)}  (p)$, $\momDD{\theta}^{(2),\RR}(p,\mu)$  and  $\lD{\theta}(\mu)$  are used as shorthands. Since  $\cwiD{\theta} (p) -  \momDD{\theta}^{(2)}(p)$ is finite 
one may use 
the same renormalisation prescription for  $\cwiD{\theta} $
\begin{equation}
\label{eq:scheme2}
 \cwiD{\theta} (p) =  \cwiR{\theta}{\theta}{\RR}(p,\mu) +  
 \lD{\theta}(\mu) \;,
\end{equation} 
connecting with the notation 
in our previous work\cite{PZ16}. 
Crucially, it is the choice (\ref{eq:scheme1}) 
of splitting the bare correlation function into a non-local renormalised part $\momDD{\theta}^{(2),\RR} $
and a local  part $\lD{\theta}$ (counterterm) which defines a scheme $\RR$ and introduces
a renormalisation scale $\mu$.\footnote{In perturbation theory the counterterm is a Laurent series 
in $\eps$ and requires the scale $\mu$. Non-perturbatively the scale $p$ is identified with $\mu$ cf. next section. 
Moreover, in what follows $\RR$ refers to the split \eqref{eq:scheme1} 
and we do not specify the renormalisation of the couplings and operators, linked by the quantum 
action principle, other than assuming a mass-independent scheme.}
The anomalous part of the equation above is
\begin{equation}
\label{eq:ZamDef}
 \zamR{\theta}{\theta}{\RR}(\mu)  = \left( \frac{d}{d \ln \mu} - 2\eps \right) \, 
 \momDD{\theta}^{(2),\RR}
  =  
   \left( \frac{d}{d \ln \mu} - 2\eps \right) \, 
  \cwiDR{\theta}{\RR}(p,\mu)
  =  
- \left( \frac{d}{d \ln \mu} - 2\eps \right) \,   \lDR{\theta}{\RR}(\mu) \;,
\end{equation}  
the quantity entering \eqref{eq:barbdmu} and 
related to the $R^2$-anomaly \cite{H81,JO90,PZ16} (eq.~48 of the 3rd reference).
The $\mu$-dependence arising through the coupling  $\zamR{\theta}{\theta}{\RR}(\mu) =  \zamR{\theta}{\theta}{\RR}(g^Q(\mu))$.   In both equations above the $\eps \to 0$ limit is smooth and we do therefore
not distinguish between a four  and $d$-dimensional $\zamR{\theta}{\theta}{\RR}$ and adapt the same attitude 
to other quantities.

\subsubsection{Definition of a \NScheme-Scheme for the $2$-point Function}
\label{sec:scheme}

Below we define a scheme which is most effectively imposed on $ \cwiD{\theta}$ rather than 
$\momDD{\theta}^{(2),\RR}  $. The renormalisation condition is 
\begin{equation}
\label{eq:mom}
\cwiDS{\theta}(p,\mu)|_{p= \mu} = 0 \;, 
\end{equation}
that the renormalised  two-point function equals zero at $ p = \mu$ (recall $p \equiv \sqrt{p^2}$) 
which is straightforwardly implemented by
\begin{equation}
\label{eq:momI}
\cwiD{\theta}(p)  =  \underbrace{\left( \cwiD{\theta}(p) -  \cwiD{\theta}(\mu) \right) }_{\cwiDS{\theta}(p,\mu)} +  \underbrace{\cwiD{\theta}(\mu)}_{\lDS{\theta}(\mu)} \;.
\end{equation}
This is equivalent to the so-called \Nscheme-scheme  (and variations thereof), 
introduced for  lattice Monte-Carlo simulations \cite{MOM},  
where the renormalised momentum space correlation function is set to its tree-level 
value for some momentum configuration set to equal $\mu$.  
A solution to eqs.~(\ref{eq:ZamDef},\ref{eq:mom}) is given by  
\begin{equation}
\cwiDS{\theta}(p,\mu)  = \int^{\ln \mu/\mu_0} _{\ln p/\mu_0}  \zamS{\theta}{\theta} (\mu')  d\ln \mu'  \;,
\end{equation}
and  therefore 
\begin{eqnarray}
\label{eq:Cscheme}
\cwiD{\theta}(p) &\;=\;& 
\underbrace{\int^{\ln \mu/\mu_0} _{\ln p/\mu_0}  \zamS{\theta}{\theta}(\mu')  d\ln \mu'}_{\cwiDS{\theta}(p,\mu)} + 
\underbrace{\int^\infty_{\ln \mu/\mu_0}  \zamS{\theta}{\theta} (\mu') d\ln \mu'+ \cwiD{\theta}(\infty)}_{\lDS{\theta}(\mu)}
 \nonumber  \\[0.1cm] 
&\;=\;&  \int^\infty_{\ln p/\mu_0}  \zamS{\theta}{\theta}(\mu')  d\ln \mu'  
+ \cwiD{\theta}(\infty)  \;.
\end{eqnarray}
Together with \eqref{eq:state} this implies eq.~\eqref{eq:barbdmu} in the \Nscheme-scheme and
 allows us to obtain $\zamS{\theta}{\theta}(\mu)$ from $\cwiD{\theta}(p)$ as follows
\begin{equation}
\label{eq:chissOS}
\zamS{\theta}{\theta}(\mu) = - \frac{d}{d \ln p} \Big|_{ p = \mu} \cwiD{\theta}(p) \;.
\end{equation}
 Since the the Lie derivative with respect to the $\be$-function vector field commutes 
 with the $\be$-functions themselves (cf. section \ref{sec:schemeD} for more details)  
\begin{equation}
\label{eq:Zamo}
\zamR{\theta}{\theta}{\RR} = \be^A \be^B \zamR{A}{B}{\RR} \;
\end{equation}
holds. Together with $p$-independence of the 
 the $\be$-functions this implies 
in the \Nscheme-scheme 
 \begin{equation}
\label{eq:chiABOS}
\zamS{A}{B}(\mu)  = - \frac{d}{d \ln p}\Big|_{ p = \mu} \cwi{A}{B}(p,\mu) \;.
\end{equation}
Above 
\begin{equation}
\label{eq:GAB}
 \Gamma_{AB}(p,\mu) =   \int d^4x e^{i p\cdot x} \vev{[O_A(x)][O_B(0)]}_c   = p^4 \cwi{A}{B}(p,\mu)   + \dots
\end{equation} 
in analogy with  \eqref{eq:4mom} where the $\mu$-depdence  comes from the the renormalisation of $[O_{A,B}]$.  
Eq.~\eqref{eq:chiABOS} is consistent with the representation of the Zamolodchikov-metric
 in conformal field theories (CFTs) 
  $\cwi{A}{B}(p,\mu) = - \zamS{A}{B}(\mu) \ln(p/\mu_0)  + const.$   
(e.g.  \cite{Prince15}) where the coupling space is referred to as a conformal manifold.  The difference is that we consider the Zamolodchikov-metric flowing between two FPs rather than in a CFT only. 
Transformation under scheme changes 
for  $\zamS{\theta}{\theta} $ and $\zamS{A}{B}$ are discussed in section \ref{sec:schemeD}. 
The formulae of this section allow us to clarify that 
 \eqref{eq:state} invariant under \eqref{eq:amb} is to be adapted to
\begin{equation}
 \label{eq:Dbarbw0}
 \Delta \barb 
 =  \frac{1}{8} ( \momDD{\theta}^{(2)}(0)- \momDD{\theta}^{(2)}(\infty) )  \;.
 \end{equation}
 In order to see this note that   \eqref{eq:CM} still holds under \eqref{eq:amb},  $\momDD{\theta}^{(2)}(\mu) \to  \momDD{\theta}^{(2)}(\mu) + \omega_0$, 
 and that in \eqref{eq:Dbarbw0} the arbitrary $\omega_0$  simply cancels in the difference on the right hand side (RHS).

\subsubsection{Positivity of the Zamolodchikov-Metric in the \Nscheme-Scheme}
\label{sec:zampos}

From the positivity of the spectral function $\rho(s) \geq 0$ and \eqref{eq:Cdisp} 
it follows that $\cwiD{\theta}(p)$  strictly increasing when $p$ decreases.  
This in turn with \eqref{eq:Cscheme} implies that 
\begin{equation}
\label{eq:Zamtt}
\zamS{\theta}{\theta}(\mu) > 0  \;\quad  \text{for } \mu \geq 0 \;.
\end{equation}
From the spectral representation of $\cwi{A}{B}$ and \eqref{eq:chiABOS} 
it follows that 
the Zamolodchikov-metric $\zamS{A}{B}$ itself,
\begin{equation}
\label{eq:ZamAB}
\zamS{A}{B}(\mu) > 0   \;\quad 
 \text{for }  \mu \geq 0    \;,
\end{equation}
is also a positive matrix along the flow.  In both cases strict positivity is 
tied to non-trivial unitary theories.
Note that even if the spectral representation of $\cwi{A}{B}$ had a logarithmic 
divergence then it would vanish under the $p$-derivative.

In 2D a positive definite Zamolodchikov-metric has been defined  
by Osborn \cite{O91} through the Weyl consistency relations and later in \cite{FK09}  
via a derivative of a configuration space cut-off.  
Our definitions seem more closely related to the latter than the former. 
We are not aware of a direct extension of the definitions in \cite{O91,FK09} to 4D.
However,  such a question has been raised  in the review \cite{N13} 
without any detailed analysis.

\subsection{A scheme for which  the $R^2$-anomaly (or $\bb$)  vanishes along the Flow}
\label{sec:R2}

The general  formalism allows us to define different schemes for different couplings by  
splitting the bare coupling into a renormalised and counterterm part.  
This applies in particular 
to gravity couplings, related to vacuum graphs,
 \begin{equation}
 \label{eq:Lgrav}
 {\cal L}_{\textrm{gravity}} = -( a_0 E_4 +   b_0  H^2 + c_0 W^2 )   \;.
 \end{equation}
 Below we define scheme for $b_0$, named \Rsscheme-scheme, for which 
 $\be_b = 0$ outside the FP  \emph{and} for which $\barb$ is governed by a gradient flow type equation. 
It is  noted that this is a priori  possible  since $\be_b = 0$ for CFTs \cite{D77,BCR83} which define the endpoints 
of the flow.
At the technical level $\be_b =0$ is established by the remarkable link between 
$\vev{\TEMTO \dots \TEMTO}$-correlators  
and the gravity terms \eqref{eq:Lgrav}   by  the QAP e.g. \cite{H81,JO90,S16}.

We find it helpful to  think of $b_0$ as the coupling of the  $R^2$-term similar to the role of the 
QCD-coupling and the field strength tensor squared $G^2$.  
Although the $R^2$-term is not  quantised itself, $b(\mu)$ runs since it mixes with other dynamical 
operators e.g. the $G^2$-term in QCD-like theories. 
The key observation is that the UV-finiteness of the fourth moment 
(or $\cwiD{\theta} (0) $) \eqref{eq:4mom} then allows to absorb this finite part 
into the renormalisation of $G^2$ in which case $\be_b =0$ along the flow. 
 
In order to make this statement transparent 
 it proves useful to briefly digress and clarify the effect 
of the choice of scheme for a coupling $g^Q$  on the conjugate renormalised composite operator $[O_Q]$.
 A choice of scheme $\Ra$ is given, as usual,  by a separation of 
the bare coupling into a renormalised coupling $ g^{Q,\Ra}(\mu)$ and counterterm $L_Q^\Ra(\mu)$
\begin{equation}
\label{eq:gsplit}
g^Q_0 = \mu^{d-4}( g^{Q,\Ra}(\mu) + L_Q^\Ra(\mu)) \;.
\end{equation}
For clarity let us mention that we have  previously  suppressed the $\Ra$-label when talking about dynamical couplings.   
The bare couplings are independent of the RG-scale, $\frac{d}{d \ln \mu} g^Q_0  =0$, and 
$L_Q^\Ra(\mu)$ therefore determines $g^{Q,\Ra}(\mu)$ up to a constant which has to be determined 
experimentally. 
The local QAP  defines the renormalised composite operator  by 
\begin{equation}
\label{eq:VEVQAP}
\vev{[O_Q(x)]^\Ra_\Rb} = ( - \de_{g^Q(x)} )\Big|^{g^A = g^{A,\Ra}}_{ v = v^\Rb}     \ln \Zpart  \;,
\end{equation}
where $v = a,b,c$ from \eqref{eq:Lgrav}  and $g^A$ are generic couplings. 
In principle one may choose different 
schemes for different couplings and parameters which leads to a proliferation of scheme dependences 
on the left hand side (LHS). 

Returning to our task we define the coupling
\begin{equation} 
\label{eq:b-coupling}
b_0 = \mu^{d-4} ( b^\RB + L_b^\RB) \;,
\end{equation}
 in analogy with \eqref{eq:gsplit} and assume 
a renormalisation scheme $\RT$ for the $\vev{\TEMTO \TEMTO}$-correlator.\footnote{We comment 
on other ways of handling the $R^2$-term in the literature in appendix \ref{app:2-ways}.} 
 A double variation of the metric  ($\gm_{\mu\nu} \to e^{ - 2 s(x)} \gm_{\mu\nu}$) is finite 
 since both the partition function and the metric are finite. 
When Fourier transformed and projected on the $p^4$-structure 
one obtains
\begin{eqnarray} 
\label{eq:2ptEM}
 \int d^d xe^{i p \cdot x} \left(  \left(- \de_{s(x)}  \right) \left(- \de_{s(0)} \right) \ln \Zpart \right)  |_{p^4} = 
   \int d^d x e^{i p \cdot x}  \vev{ \TEMTO(x)  \TEMTO(0)}|_{p^4}+ 8  \, b_0  =  \fin \;.
\end{eqnarray} 
This implies the non-trivial, known,  relation 
\begin{equation}
\label{eq:link}
L_b^{\RB}  =- \frac{1}{8}\lDR{\theta}{\RT} + \fin  \;,
\end{equation}
quoted for in the $\MS$-scheme in \cite{F83}.
The difference in signs in \eqref{eq:link} is 
somewhat unfortunate but imposes itself in this sector cf.  \cite{PZ16} for more detailed remarks.

The observation that the finiteness of $\lDR{\theta}{\RT} $ implies the finiteness of  $L_b^{\RB}$ 
can be used to define a scheme, which we call \Rsscheme-scheme, 
$b_0 =  \mu^{d-4}   ( b^\Rs + L_b^\Rs)$ with
\begin{equation}
\label{eq:R2-scheme}
b^\Rs = b + L_b \;, \quad  L_b^\Rs = 0 \;.
\end{equation}
This is equivalent to saying that it is not necessary to renormalise since there are no divergences. In the \Rsscheme-scheme we therefore have that 
\begin{equation}
\label{eq:beb}
\be_b^\Rs(\mu) = - (\frac{d}{d \ln \mu}  - 2 \eps) L_b^\Rs   =   0 \;, \quad \mu \geq 0 \;.
\end{equation}
This means  that  the $b^\Rs$-coupling does not receive RG-running
 by other dynamical operators.\footnote{Where the characterisation ``other" refers to the fact that  
the $R^2$-gravity term is not quantised and therefore does not contribute to the running 
of the $b^\Rs$-coupling. Whether or not in such a case a scheme  exist where the  
$R^2$-coupling does not run is beyond the scope of investigations of this work. 
This question can be posed in a well-defined framework, modulo ghosts due to higher derivatives, 
 since  $R^2$-gravity has been shown 
to be renormalisable  \cite{Stelle}.}
All that remains is to determine the previously mentioned  unknown constant by experiment. 
The VEV of the TEMT, $\vev{\TEMT{\rho}}$, is of course  invariant under scheme-changes as  
illustrated in section \ref{sec:R2QCD} for QCD-like theories.

Before continuing towards the flow of $\Box R$-term 
we  digress in discussing whether  or not  schemes could exist 
for which the other Weyl-anomaly  \eqref{eq:VEVTEMT} vanish along the flow. 
An a priori no-go argument is 
that, unlike the $R^2$-anomaly,  the other anomalies  have generically a non-zero flow difference.
We consider two types of  gravitational trace anomalies  (cf. \cite{DS93} for a more refined discussion without 
inclusion of $\Box R$ though):
\begin{itemize}
\item $\be$-functions terms. For the $\be_{a,c}$-function terms, the analogous argument 
as above would require $L_a$ and $L_c$ to be finite.  
\begin{itemize}
\item $\ba E_4$-term:  The counterterm of $E_4$ has been shown to be finite only 
when mulitiplied by $\eps$ \cite{PZprep17}. This is  typical  for 
topological terms since their non-total derivative parts are necessarily evanescent. 
The local QAP then implies finiteness constraints on $\eps L_x$ 
where $L_x$ is the counterterm associated with the topological invariant.
Since $L_a$ is not finite we conclude that there does not exist a scheme where 
$\ba$  can be set to zero along the flow.
\item $\bc W^2$-term: The $W^2$ term is associated with the spin $2$ part of the 
$\vev{\TEMTO_{\rho \sigma} \TEMTO_{\la \nu} }$-correlator. The latter is generically divergent in the relevant structure  
contrary to 
 the $ \vev{\TEMTO \TEMTO }$-correlator. The essential point is that the TEMT is protected in the UV by the additional couplings originating from 
the dynamical $\be$-functions. For example in QCD-like theories 
$\TEMTO \sim \be G^2 + \dots $  whereas $\TEMTO_{\rho \sigma} =      \frac{1}{4} g_{\rho \sigma} G^2  - G_{\rho \al} G^{\al}_{\phantom{\al} \sigma} + \dots $. In the convergence criterium for asymptotically free theories
in \cite{PZ16}, this means that $n_{\TEMTO\TEMTO} = 2$ and $n_{\TEMTO_{\rho \sigma} \TEMTO_{\la \nu}} = 0$ which satisfies and violates the convergence criteria in section 
3.1 of this reference. Hence we conclude that $L_c$ is not finite when the regulator is removed and $\bc$ can therefore not be set to zero.
\end{itemize}
\item $\barb  \Box R$-term: Is not a $\be$-function term and therefore does not derive 
from \eqref{eq:beb}. Thus the same trick is not applicable.
\end{itemize}

\subsection{Properties of $\Delta \barb$,  $\barb^\RT_\RB(\mu)$ and the Zamolodchikov-metric $\zam{A}{B}^\RT$}
\label{sec:prop}

Clarifying the properties of the quantities $\Delta \barb$,   $\barb^\RT_\RB(\mu)$  and $\zam{A}{B}^\RT$ 
is linked to  understanding their scheme dependences. The following hierarchy or degree of complication emerges.
 The global flow 
$\Delta \barb$ (section \ref{sec:global}) is scheme-independent. 
The local flow properties, discussed in section \ref{sec:local}, are scheme-dependent. 
The infinitesimal change along  the flow $\frac{d}{d \ln \mu}  \barb^\RT
=   \frac{1}{8}    \zamR{A}{B}{\RT} \be^A \be^B  $ \eqref{eq:gf} is dependent on the $\RT$-scheme and 
the local value $\barb^\RT_\RB(\mu)$  is dependent on both the $\RT$- and $\RB$-scheme.

\subsubsection{Properties of $\Delta \barb$ (global Flow)}
\label{sec:global}

Let us summarise the various ways in which $\Delta \barb$ \eqref{eq:state} can be expressed as an integral  
using \eqref{eq:mom4D}, \eqref{eq:Cdisp} and  \eqref{eq:Cscheme}\footnote{Formally 
 $\vev{ \TEMTO(x) \TEMTO(0)}_c^{\omega_0}  =  \vev{ \TEMTO(x) \TEMTO(0)}_c- \omega_0 \Box^2 \de(x)$
 where $\vev{ \TEMTO(x) \TEMTO(0)}_c$ is 
 evaluated 
 by any regulator respecting the symmetries and $\omega_0 =  \momDD{\theta}^{(2)} (\infty)$  is 
 assumed for definiteness.
The regulator $\RR$ can be removed smoothly since the moment is UV-finite.}
 \begin{eqnarray}
\Delta \barb \; =\; 
 \frac{1}{8} \left(  \momDD{\theta}^{(2)} (0) -   \momDD{\theta}^{(2)}(\infty) \right) 
  \label{eq:barb0}   &\; =\; &    \frac{1}{2^9\, 3} 
\int d^4 x\, x^4   \vev{ \TEMTO(x) \TEMTO(0)}^{\omega_0}_c  
 \\[0.1cm]
 \label{eq:barb2}
 &\; =\; & \frac{1}{8}  \int^\infty_{-\infty}  \zamR{\theta}{\theta}{\RR} ( \mu')  d\ln \mu'    \\[0.1cm]
 \label{eq:barb1}
  &\; =\; &  \frac{1}{8} \int_0^\infty d s \frac{\rho(s)}{s^{3}} 
       > 0
  \;.
\end{eqnarray}
 The following properties are immediate 
\begin{itemize}
\item \emph{Positivity:} $\Delta \barb > 0$ follows from the positivity of the spectral function 
$\rho(s) \geq 0$ as well as the positivity of the Zamolodchikov-metric  in the  \Nscheme-scheme \eqref{eq:ZamAB}. Since  $\Theta_{\textrm{CFT}} \to 0$ and therefore 
$\Delta \barb|_{\textrm{CFT}} = 0$, a non-zero value 
 measures the departure from  conformality. 
Note that  the  $\TEMTO(x)\TEMTO(0)$-correlator can be interpreted  as a probe 
that records a response of a theory with couplings $g^A(\mu = x^{-1})$.

\item  \emph{Scheme-independence} of $\Delta \barb$ follows from the scheme-independence 
of the spectral function $\rho(s)$ and the fact that the spectral representation does 
not require subtractions. Similarly since $\Delta \barb$ can be expressed in terms 
of a bare correlation function \eqref{eq:barb0} the scheme-independence of the latter implies  scheme-independence of $\Delta \barb$. Further remarks on scheme dependence and independence can be 
found in section \ref{sec:schemeD}.

\item \emph{Flow-independence} follows from combing  eqs.~\eqref{eq:ZamDef} and \eqref{eq:gf}  into 
\begin{equation}
\label{eq:combine}
\frac{d}{d \ln \mu}  \barb^\RT = \frac{1}{8} \frac{d}{d \ln \mu}   \cwiR{\theta}{\theta}{\RT}(p,\mu)  \;,
\end{equation}
which shows that the flow of $ \barb^\RT$ derives from a potential and is therefore independent 
of the flow itself.
More explicitly this equation, when  integrated  over $d\ln \mu$ and particularised to 
the \Nscheme-scheme,  gives
\begin{eqnarray}
\label{eq:flowI}
\Delta \barb  &\;=\;&  
\frac{1}{8} \left( \cwiR{\theta}{\theta}{\MOM}(p,\infty)  -   \cwiR{\theta}{\theta}{\MOM}(p,0)  \right) 
\nonumber \\[0.1cm]
&\;\stackrel{\eqref{eq:momI}}{=}\;&    \frac{1}{8} ( \cwiD{\theta}(0)- \cwiD{\theta}(\infty) )
\nonumber \\[0.1cm]
&\;\stackrel{\eqref{eq:CM}}{=}\;&  \frac{1}{8} \left(  \momDD{\theta}^{(2)} (0) -   \momDD{\theta}^{(2)}(\infty) \right) \;,
\end{eqnarray}
equation \eqref{eq:Dbarbw0}. 
Hence this derivation provides an alternative  to the one  presented in section \ref{sec:scheme}.
Flow-independence only poses itself for two or more couplings, as illustrated in fig.~\ref{fig:flow}, 
and translates in our case to the question whether (the difference of) $2$-point functions 
can depend on the approach in coupling space.  Local  reversibility of  RG-flows  
implies that this cannot be the case. If one assumes for example that the RG-flow can be linearised 
around a FP then the limit is automatically uniform and the flow therefore independent of the path.
\begin{figure}[h]
\centering
\includegraphics[width=9cm,clip]{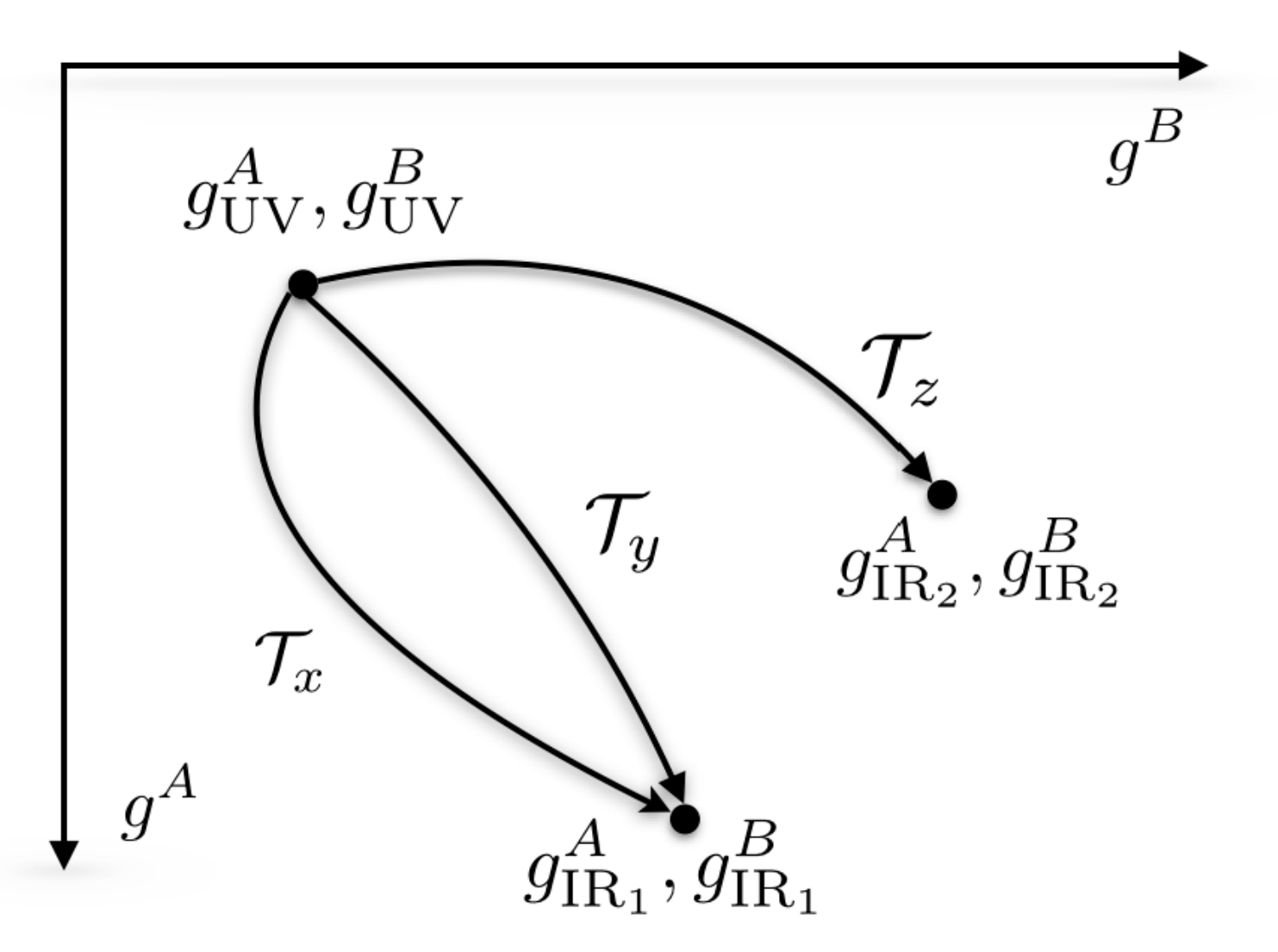}
\caption{\small Possible RG-flow trajectories from an UV-FP $g^A_\UV,g^B_\UV$ to an IR-FPs. 
The trajectories ${\cal T}_{x,y}$ and ${\cal T}_z$ flow into the 
 IR$_1$- and  IR$_2$-FPs respectively.
Hence $\Delta \barb_{{\cal T}_x} = \Delta \barb_{{\cal T}_y}\neq \Delta \barb_{{\cal T}_z}$ with the last statement being the generic case.}
\label{fig:flow}       
\end{figure}

Equivalently flow-independence can  be obtained by rewriting  \eqref{eq:barb2} as line integral 
of a vector $V_B^\RR  $ over  coupling space 
\begin{eqnarray}
\label{eq:stokes}
\Delta \barb  =  \frac{1}{8} \int^\infty_{-\infty}  \be^A \be^B \zamR{A}{B}{\RR} d\ln \mu' =\frac{1}{8} \int_{\vec{g}_\IR }^{\vec{g}_\UV } V_B^\RR dg^B \;.
\end{eqnarray}
Path-independence follows from 
$V_B^\RR$ being curl-free which is  true if and only if $V_B^\RR$ derives from a potential 
$V_B^\RR =  - \partial_B f^\RR $. Contracted by $\be^B$ gives 
$\zamR{\theta}{\theta}{\RR}  =   \be^B V_B^\RR =  - \be^B \partial_B f^\RR = - \frac{d}{d \ln \mu }f^\RR $
 for which 
$f^\RR  =  \lDR{\theta}{\RR}$ is a solution \eqref{eq:ZamDef}.  We refer the reader to appendix \ref{app:asym} 
for related and refined discussion of these quantities. 
Note that we have used that $\lDR{\theta}{\RR}$ is independent of $b$ in writing $\be^B \partial_B  \lDR{\theta}{\RR} $ 
as a total $\ln \mu$-derivative of  $ \lDR{\theta}{\RR}$. 

It should be added that flow-independence is not 
straightforward in the case where the coupling manifold 
is topologically non-trivial e.g. not simply connected. 
In this case the Stokes like argumentation \eqref{eq:stokes} 
breaks down and the correlation functions in \eqref{eq:flowI} are 
multivalued. This topic certainly deserves further study but is 
beyond the scope of this paper and we refer the reader to 
Ref.~\cite{Gukov:2015qea} for recent discussion on how to count RG-flows. 
\end{itemize}

\subsubsection{Properties of $\barb(\mu)= \barb^\RT_\RB(\mu)$ outside the Fixed Points (local Flow)}
\label{sec:local}

The extension of $\barb$ outside the FP is scheme-dependent. 
It is dependent on the scheme for the $\vev{\TEMTO\TEMTO}$-correlator and 
the $b$-coupling which were discussed in sections \ref{sec:Gscheme} and \ref{sec:R2} respectively.  
Hence generically  $\barb(\mu) = \barb^\RT_\RB(\mu)$.
For extending the flow integral the preferred 
 scheme is the \Nscheme-scheme  where 
the Zamolodchikov metric is positive and properties of monotonicity and gradient flow follow. 
\begin{itemize}
\item 
\emph{Monotonicity:}  From \eqref{eq:Cscheme} we may define,  
\begin{equation}
\label{eq:mono}
 \barb^\MOM_{\RB}(\mu) 
=    \barb^\UV_{\RB} -  \frac{1}{8} \int^\infty_{\ln \mu/\mu_0}  \zamS{\theta}{\theta}(\mu') d\ln \mu' \;,
\end{equation}
a flow dependent extension  satisfying the boundary conditions $ \barb^\MOM_{\RB}(\infty) =  \barb^\UV_\RB $ 
and  $ \barb^\MOM_{\RB}(0) = \barb^\IR_\RB$. Due to the positivity of $\zamS{\theta}{\theta}$ \eqref{eq:Zamtt} 
the function $ \barb^\MOM_{\RB}(\mu)  $ is monotonically decreasing along the flow (with decreasing $\mu$).
\item \emph{Gradient flow type equation:}  From the anomalous WI \eqref{eq:AWI} 
the following equation \eqref{eq:gf} was derived\,\footnote{Note that the $\RT$ scheme-dependence of the Zamolodchikov-metric and $\barb$ ought to cancel 
on the RHS of the second equation.
Eq.~\eqref{eq:gfp} is equivalent to one of Osborn's Weyl consistency relations  cf. eq.~3.10c in \cite{O91} 
upon identifying  $\zamR{A}{B}{\MS} \to -\zam{A}{B}^a$ and $4 \barb^\MS \to  d + \frac{1}{2}U_I \be^I$.} 
\begin{equation}
\label{eq:gfp} 
   \frac{1}{8}    \zamR{A}{B}{\RT} \be^A \be^B  =  \frac{d}{d \ln \mu}  \barb^\RT = 
  (  \be^A \partial_A  + \be^\RB_b \partial_b) \barb^\RT_\RB  \;.
\end{equation}
For $\RT = \;$\Nscheme-scheme, \eqref{eq:gfp} would be a gradient flow type equation if 
it were not for the  $\be_b^\RB$-term. Since the latter vanishes in the $\RB=$\;\Rsscheme-scheme one can then obtain a gradient flow 
equation. 
For a compact presentation of the gradient flow formulae the 
following notation is introduced 
\begin{equation}
\label{eq:short}
\barbb(\mu) \equiv  \barb^\MOM_{\rsscheme}(\mu)  \;, \qquad \GM_{AB}(\mu) \equiv  \frac{1}{8} \zamR{A}{B}{\MOM}(\mu) \;,
\end{equation}
and $T = - \ln \mu$, increasing towards the IR, and   shorthand $\dot{\phantom{x}} = \frac{d}{dT}$. 
The equation then assumes the familiar form
\begin{equation}
\label{eq:gfpp}
 \dot{\barbb} = - \be^A \partial_A   \barbb  = -  \GM_{AB} \be^A \be^B <  0 \;.
\end{equation}
One then obtains the gradient flow type equation of  the form
\begin{equation}
\label{eq:GF}
\partial_A   \barbb =      ( \GM_{AB}  + \AS_{AB} )  \be^B  \;, 
\end{equation}
where $ \AS_{AB} = - \AS_{BA}$ is an antisymmetric part whose form 
is discussed in appendix \ref{app:asym}. In the case where the antisymmetric part vanishes, 
\eqref{eq:GF} becomes a proper gradient flow equation and
can be inverted to give  $\be^B =  \GM^{AB}  \partial_A   \barbb$ 
where $\GM^{AB} \equiv  (\GM_{AB} )^{-1}$ is the inverse matrix which exists since the eigenvalues
of $\GM^{AB}$ are strictly positive \eqref{eq:ZamAB}.
 Covariance of  equation \eqref{eq:GF} under couplings scheme change is shown in the next section. Note that eq.~\eqref{eq:GF} is though  not covariant under $\RB$-scheme changes.
\end{itemize}

\subsubsection{Transformation of the  Zamolodchikov-Metric under Scheme changes}
 \label{sec:schemeD}
 
 The Zamolodchikov-metric  has been implicitly defined through  \eqref{eq:ZamDef} and \eqref{eq:Zamo} 
 in an arbitrary scheme $\RT$ and explicitly for the \Nscheme-scheme \eqref{eq:chiABOS}. 
  The expression of $\Delta \barb$ \eqref{eq:barb0} is obviously scheme independent and so the question of how the scheme dependence of $\zamS{\theta}{\theta}$ cancels in 
  the representation \eqref{eq:barb2} is of interest
  which we aim to clarify in this section.
 It is appropriate to distinguish between the scheme dependence due to 
 renormalisation condition \eqref{eq:mom}, denoted by $\RT$,  and  a redefinition of the $g^Q$-scheme of the dynamical couplings \eqref{eq:gsplit} which we have ignored for most part of the paper. 
 It is worthwhile to emphasise  that the transformations have a geometric interpretation in the space of couplings 
 in that  the $\RT$-transformation is governed by a Lie derivative on 
 a $2$-tensor (infinitesimal change of a tensor along a flow) 
 and the $g^Q$-transformation  corresponds to a coordinate change (generalised rotation).
 \begin{enumerate}
 \item Changing the renormalisation from $\RT_1$ to  $\RT_2$ corresponds to 
  \begin{equation}
 \label{eq:scheme-change}
  \momT{AB}^{\RT_2} =  \momT{AB}^{\RT_1} + \omega_{AB} \;,  \quad 
    \lnDR{A}{B}{\RT_2} =  \lnDR{A}{B}{\RT_1} - \omega_{AB} \;,
 \end{equation} 
 where $\omega_{AB}$ is finite, local and  $\mu$-dependent.  
The split  $\momT{AB} (p,\mu) =  \momT{AB}^{\RR}  (p,\mu)+  
  \lnD{A}{B}(\mu)$ is defined in analogy to   \eqref{eq:scheme1}
  with regards to  the $\vev{O_A O_B}$-correlator \eqref{eq:GAB}.
With \eqref{eq:ZamDef} and \eqref{eq:Zamo} this results in
 \begin{alignat}{5}
 \label{eq:Rchange}
 &  \de \zam{\theta}{\theta}   &\;=\;&  \zamR{\theta}{\theta}{\RT_2}  -  \zamR{\theta}{\theta}{\RT_1}  &\;=\;&  
 {\cal L}_\be \omega  &\;=\;&     \be^Q \partial_Q  \omega  &\;=\;&  
      \frac{d}{d \ln \mu} \omega  \;, \nonumber \\[0.1cm]
 &   \de \zam{A}{B}  &\;=\;& \zamR{A}{B}{\RT_2} -  \zamR{A}{B}{\RT_1}  &\;=\;&   
    {\cal L}_\be  \omega_{AB}   &\;=\;&     \be^Q \partial_Q \omega_{AB} &\;+\;&   
\left\{ (\partial_B \be^Q )\omega_{AQ} + {A \leftrightarrow B} \right\}   \;,
  \end{alignat}
  where the abbreviation $\omega = \be^A \be^B \omega_{AB} $ was used and 
  ${\cal L}_\be$ denotes the Lie derivative with respect to the vector field $\be^A$.
  Hence \eqref{eq:barb2} is manifestly invariant under the scheme change 
\eqref{eq:scheme-change} provided $\omega$ vanishes at both the UV and IR 
boundary. 
Under such circumstances a scheme change might be regarded
as being cohomologically trivial.
 Incidentally \eqref{eq:Rchange} also clarifies that the 
Zamolodchikov metric for a scheme, other than \Nscheme-scheme, is defined as follows\footnote{Eq.\eqref{eq:zamR} can either be derived by straightforward computation 
or one may use that on a scalar, with no explicit $\mu$-dependence, 
 $\frac{d}{d \ln \mu}=  \be^C \partial_C = {\cal L}_\be$ and that the Lie derivative 
  along a vector field acts trivially on itself. 
The reason that the general definition is more involved, than the \Nscheme-scheme, \eqref{eq:chiABOS} is that the  $\mu$-, unlike the $p$-, derivative does not commute 
with the $\be$-functions.}
\begin{equation}
\label{eq:zamR}
 \zamR{A}{B}{\RR} = - {\cal L}_\be   \lnD{A}{B}(\mu)   =
- \left(  \be^Q \partial_Q   \lnD{A}{B}(\mu)  + 
\left\{ (\partial_A \be^Q)   \lnD{Q}{B}(\mu) + {A \leftrightarrow B} \right\}  \right)
 \;.
\end{equation}
 It is noteworthy that this does not correspond to  a total derivative with respect to $ \ln \mu$.

 \item Independence under a change in the coupling constant scheme follows
 from the $\be$-function as well as  $\zamS{A}{B} $ transforming as tensors. 
 Going from the scheme $g^P \to g^{'P}$ results in 
  \begin{equation}
 \be^{'P} =    \frac{\de g^{'P}}{\de g^A}  \be^A  \;,  \quad 
 \zamP{P}{Q} = \frac{\de g^A}{\de g^{'P}}\frac{\de g^B}{\de g^{'Q}}  \zamS{A}{B} \;,
 \end{equation}
 where the first equation results from the chain rule and so does the second 
 since $\zamS{A}{B}$ is derived from
 \begin{eqnarray}
  \vev{[O_A(x)] [O_B(0)] }_c' &\;=\;& (- \de_{'A(x)})(-\de_{'B(0)}) \ln \Zpart  = 
  \frac{\de g^P}{\de g^{'A}}\frac{\de g^Q}{\de g^{'B}}  
  (- \de_{P(x)})(-\de_{Q(0)}) \ln \Zpart  \nonumber \\[0.1cm]
   &\;=\;&   \frac{\de g^P}{\de g^{'A}}\frac{\de g^Q}{\de g^{'B}}   \vev{[O_P(x)] [O_Q(0)] }_c  \;,
 \end{eqnarray} 
 where the prime denotes the change of the coupling scheme and 
 $\de_{'A(x)} = \de/\de g^{'A}(x)$.
  Clearly 
 $\be^{'P} \be^{'Q}  \zamP{P}{Q} = \be^{A}\be^{B}   \zam{A}{B}$ which shows the scheme
 independence.
 \end{enumerate}

\subsection{UV and IR Convergence the $\Delta \barb$-Integral Representation}
\label{sec:convergence}

For eqs.~(\ref{eq:mom4D},\ref{eq:barb1},\ref{eq:barb2}) being a valid way to compute 
$\Delta \barb$ the integrals need to be finite.  We shall see that \eqref{eq:barb1} is not finite
in the spontaneously broken phase which implies that either $\Delta \barb$ diverges or that
the formalism needs to be adapted.
Before investigating the representation  \eqref{eq:barb1} 
it is instructive to consider  
$\Delta \barb \sim \int d^4 x\, x^4   \vev{ \TEMTO(x) \TEMTO(0)}_c$ \eqref{eq:mom4D}. 
Firstly, $\vev{ \TEMTO}$ 
is well-defined since $\vev{\TEMT{\rho}} $ is scale independent and differs from 
$\vev{ \TEMTO}$ by the finite Weyl-anomalies vanishing in flat space. 
Hence it is the correlation of the two  $\TEMTO$-operators which is subject to potential divergences 
in the UV ($x \to 0$) as well as the IR (for $x \to \infty$). 

The technical discussion parallels the one in \cite{LPR12} 
with a slightly more refined discussion on the subtle case of the 
chirally broken phase in section \ref{sec:chiral}.
In order to analyse the UV- and IR-convergence one needs to investigate the behaviour 
of the spectral function close to the FP. 
In the case where  the scaling dimension (i.e. classical plus anomalous dimension) of the most relevant operator is $\Delta $ 
the spectral density \eqref{eq:spectral} behaves like $\rho(s) \sim s^{\Delta -2}$ and from \eqref{eq:barb1}
\begin{equation}
\label{eq:disp2}
\Delta \barb   \sim   \int_0^\infty \frac{d s}{s}   s^{\Delta -4} \;.
\end{equation}
It is understood that the identity operator (i.e. the cosmological constant term), for which $\Delta = 0$, 
is subtracted by an appropriate  UV-counterterm as otherwise $\rho(s) \sim s^{-2}$. 

It is useful to distinguish the cases of a non-trivial  and a trivial FP.
(i.e. asymptotically safe (AS) and asymptotically free (AF)).
The case where there is spontaneous breaking of chiral symmetry is subtle 
cf.~section \ref{sec:chiral}.
For the AS-case $\Delta_{\UV} > 4 $ and $\Delta_{\IR} < 4$ in which  case 
the dispersion representation \eqref{eq:disp2} converges both in the UV and the IR.
For the AF-case $\Delta = 4$ \eqref{eq:disp2}  is potentially both divergent in the 
IR and UV  requiring a refined discussion taking into account the logarithmic behaviour.
In our previous paper \cite{PZ16} it was shown that AF-free theories, 
including the multiple coupling case, converges in the UV. 
In perturbation theory this can be seen by resumming the logarithms order by order.
An IR-AF theory  behaves in the same way with $s \to s^{-1}$ which leads 
to the same integral as in the UV \cite{LPR12} and is therefore convergent. 

In conclusion in all cases where the theory is a CFT in the IR and UV  the integral 
representations \eqref{eq:barb1},\eqref{eq:barb2}  and \eqref{eq:mom4D} are finite and do hold.
Potential problems with the formulae occur when the theory is not 
a CFT in either the UV or IR.  This is not surprising since for 
the IR effective action derivation of \eqref{eq:mom4D} (cf. appendix \ref{app:athm}),
conformality at the FPs is an assumption.
The cases where the FPs are not conformal include the free massive non-conformally
coupled scalar and  the free massive vector boson (cf. section \ref{sec:free}), 
as well as the  chirally broken case which might belong to the former type in the  IR.  
A few short comments on extending the framework to include dimensionful couplings. 
Generally dimensionful couplings should not worsen the UV-convergence. For example 
applying the fourth moment projector  $\hat{P}_2$ to the fermion 
correlator $m^2 \vev{\bar qq \bar qq}$,  in appendix B in \cite{PZ16}, 
the $p \to \infty$ limit exists ensuring UV-finiteness. The convergence in the IR is less obvious
but if the dimensionful parameter is a mass the latter can act as an IR cut-off and is therefore 
expected to improve the IR-behaviour.

\subsubsection{Spontaneous broken Chiral Symmetry in the Infrared}
\label{sec:chiral}

The case of  spontaneously broken chiral symmetry (e.g. QCD) is more cumbersome  when viewed 
from standard chiral perturbation theory.   The $\pi$ goldstone bosons are free scalars in the 
far IR and the operator-part of the TEMT contains a term
$\TEMTO  = - \frac{1}{2} \Box \pi^2  + ..$ at the classical level (e.g. \cite{LS89}).
This EMT cannot 
undergo the improvement proposed in  \cite{CCJ70} which removes the term above, 
since the improvement term is incompatible with 
chiral symmetry 
\cite{DV82,LS89,DL91,LPR12}.  This is reflected in the generally accepted view that chiral symmetry and conformal symmetry are incompatible with each other. 

Adapting the view that chiral symmetry is not compatible with conformal symmetries 
may lead to problems since in this case $\be_b^\IR \neq 0$ and the $\Delta \barb$ formulae 
might need to be reconsidered. 
 The most concrete way is to approach the problem by computation.
 In the limit of free pions the 
$\vev{\TEMTO(x) \TEMTO(0)} \to \frac{1}{4} \vev{\Box \pi^2(x) \Box \pi^2(0)}$ correlator corresponds to  a bubble graph
which contributes  a term of the form  $\Gamma_{\theta\theta}(p)  \sim p^4 \ln (4 m_\pi^2 + p^2) + \dots$  
to the TEMT-correlator  where a  quark mass $m_q$ ($m_\pi^2 \sim m_q \Lambda_{\textrm{QCD}}$) was introduced as an IR-regulator. 
(cf. the closely related discussion in and around 
  eq~2.26 in  \cite{LPR12}).
This leads to $\momDD{\theta}^{(2)}(0) \sim \ln(4 m_\pi^2) + \dots$ which diverges  in the chiral limit 
$m_q \to 0$. 
 Unlike in the UV-case it does not seem possible that this behaviour is 
improved by resumming interactions since corrections necessarily come with additional powers of 
$p^2/f_\pi^2$ where $f_\pi$ is the pion decay constant. A series of the form $ \ln(4 m_\pi^2 + p^2) \sum_{n \geq 0} x_n (p^2/f_\pi^2 )^n 
(\ln(4 m_\pi^2 + p^2) )^{a_n}$ with $a_n \leq n$ does not resum to an expression which is finite in
the limit $p^2,m_\pi^2 \to 0$. This is the case  since each 
 coefficient $n \geq 1$  vanishes in this limit and the non-zero $x_0$-term gives rise 
 to a divergence.\footnote{\label{foot:practical} In principle $\Gamma_{\theta\theta}(p)  \sim p^4 \ln (4 m_\pi^2 + p^2)+ ...$ might also affect the formula for  $\Delta \be_a$ when expressed as a dispersion relation of
  the four-dilaton scattering amplitude \cite{KS11,K11}. Note that the four dilaton scattering amplitude 
 contains a term proportional to $\Gamma_{\theta\theta}(p)$, where two dilatons couple to the same TEMT 
 on each side, e.g.  \cite{LPR12,BKZR14} eq.~3.7 in the first reference. 
 This term does not vanish when the individual dilatons are put on shell since the $p^2$ variable corresponds  to the sum of two dilaton momenta $p^2= (p_1+p_2)^2$.
Whether or not such a divergence is cancelled by other terms deserves some further study.  
Clearly it is at most the formula and not the $a$-theorem itself which is troubled by the chiral phase. 
Due to the topological nature of the Euler term $\be_a$ is well-defined at each 
end. Therefore one may introduce a mass for the quarks 
and compute $\Delta \be_a$ via a two-step process 
$\Delta \be_a = \Delta \be_a|_{m_q \neq 0 } - \Delta \be_a|_{N_f^2-1 \text{ free scalars}} $ in order to take into account the $N_f^2-1$ free massless goldstone bosons.}
Hence if $\TEMTO \to - \frac{1}{2} \Box \pi^2$ is the correct prescription for a chirally broken theory 
then this implies that  $\Delta \barb$ diverges or that the formula \eqref{eq:barb1} has to be amended. 
Whether or not this prescription is really correct is not known  to our knowledge in the sense 
of being  verified by experiment. 

Hence the caveat to the reasoning above is that we do not know for sure whether 
the chirally broken phase is a  CFT in the IR or not.  
The degrees of freedom of an effective theory are  not always necessarily clear a priori or 
simply a working assumption justified a posteriori  by their success.
Low energy QCD is  described by an effective theory of pions, 
known as chiral perturbation theory ($\chi \text{PT}$), which is 
extremely successful in many domains but whether or not IR conformality per se has been tested is unclear. 
For example it has been advocated \cite{CT13} that  to describe three-flavour $\chi \text{PT}_3$  
 it might be advantageous to supplement $\chi \text{PT}$  with an additional pseudo-goldstone (dilaton)  
resulting from the spontaneous (anomalous) breaking of  scale invariance. 
The effective theory is  known as $\chi \text{PT}_\sigma$ \cite{CT13} and it is currently unclear whether or not 
this is a valid  description in the sense of improved convergence over $\chi \text{PT}_3$.
The EMT  undergoes an improvement in the dilaton field, 
which is not constrained by chiral symmetry breaking, and seems to eliminate some of 
 the dangerous kinetic terms (cf. eq.~3.7 in \cite{CT15})
discussed above.  The remaining kinetic terms are absent in the case where
the low energy constants 
$c_{1,2}(\mu) \to 1$ for $m_q,\mu \to 0$ 
which is the chiral-scale limit advocated in \cite{CT15}.
In summary in $\chi \text{PT}_\sigma$ the EMT can be improved in the dilaton sector which in principle 
allows for the elimination of the previously discussed and dangerous 
$\Box \pi^2$-term.
 It would be interesting to compute \eqref{eq:mom4D} non-perturbatively 
on the lattice and to check whether or not a chiral logarithm of the form  
$\ln m_\pi^2  \sim \ln m_q$ is present.

\subsection{Section Summary}
\label{sec:summary}

Since this section is the heart-piece of this work we summarise before continuing the paper. 
The integral representations eqs.~(\ref{eq:mom4D},\ref{eq:barb1},\ref{eq:barb2})  are well-defined when 
the theory is  conformal in the IR and UV. The latter might not  be the case 
for the chirially broken IR-phase (cf. section \ref{sec:chiral}) and 
the free field theories of the non-conformally coupled scalar and vector particle (cf. section \ref{sec:free}). 
For the latter two cases the operator-part of TEMT, which excludes 
equation of motion terms, reads $\TEMTO = - \frac{1}{2} \Box \phi^2$ 
and $\TEMTO = - \frac{1}{2}  \Box  A_\nu^2  $ which are only scale but not conformal invariant and  
the $\Delta \barb$-integral \eqref{eq:mom4D} diverges in the IR and UV respectively. 
The IR and UV divergences of the free field correlators also seem 
to be the underlying reason why these cases are found to be regularisation 
dependent in actual calculation \cite{D77,BC77,AGS05,VFGS15,Chu16} 
as documented in the classic textbook of Birrell and Davies \cite{birrell1982quantum}. 
In summary if  \eqref{eq:barb0} is well-defined then  positivity and  scheme-independence 
of the spectral function imply the global properties $\Delta \barb\geq 0$ and $\Delta \barb$ scheme-independence. 
Flow-independence follows from the fact that the integrand of \eqref{eq:barb2} can be written 
as a total $\ln \mu$ derivative \eqref{eq:ZamDef}  (with $\eps \to 0$-limit implied).
The extension of $\barb(\mu)$ outside the FP is scheme-dependent. 
In the \Nscheme-scheme (cf. section \ref{sec:scheme}) for the $\vev{\TEMTO\TEMTO}$-correlator,  
positivity of the  
Zamolodchikov-metric, as derived in section \ref{sec:zampos}, allows us to extend the $\barb(\mu)$ as a monotonically 
decreasing function \eqref{eq:mono}  and in the \Rsscheme-scheme (cf. section \ref{sec:R2}) for the $b_0 R^2$-term, $\barb(\mu)$ is shown to satisfy a gradient flow type equation \eqref{eq:gfpp}.

\section{Examples in QCD-like and Free Field Theories}
\label{sec:QCD-like}

Below details on renormalisation are illustrated in sections \ref{sec:schemeC} and 
\ref{sec:R2QCD} for QCD-like theories and examples are given for 
a CBZ-FP and free field theories in sections \ref{sec:BZ} and \ref{sec:free} respectively. 
Other examples, such as the ${\cal O}(N)$ sigma model in the large $N$ limit, 
can be found in  the earlier work \cite{Cappelli2}. This reference also discusses 
examples in $d= 4 - \eps$ and $d=3$ dimensions which are not directly related to 
our work since we strictly adapt  $d=4$ in association with the $\Box R$-flow.

\subsection{Zamolodchikov-Metric in the \Nscheme- and $\MS$-Scheme} 
\label{sec:schemeC} 

In this section we exemplify the Zamolodchikov-metric in QCD-like theories 
in the \Nscheme-scheme and the $\MS$. 
The result can be extracted to NNLO using a recent computation 
of field-strength correlator in  \cite{Zoller:2016iam}. 
The convention for the QCD coupling and  the logarithmic $\be$-function are given in 
 appendix \ref{app:QCD-like}. With these definitions the operator-part of the TEMT reads 
$\TEMTO = \frac{\be}{2} [G^2]$ and therefore $\zam{\theta}{\theta} = \frac{1}{4} \be^2 \zam{g}{g}$.

The \Nscheme-metric is obtained by using  \eqref{eq:chissOS} 
and identifying $\cwiD{g}(\als(p)) = 16 C_0^{GG}$ 
in [Eq 4.18] \cite{Zoller:2016iam} 
\begin{eqnarray}
\zamS{g}{g} &=& - \frac{d}{d \ln p}\Big|_{ p = \mu} \cwiD{g}(\als(p))  \nonumber \\[0.1cm] 
&=&  \frac{n_g}{2 \pi^2}\left(1 + \als \left( \frac{73}{3} C_A - \frac{28}{3} N_F T_F \right) \right)   + {\cal O}(\alsp{2})  \;.
\end{eqnarray}
with $\als \equiv g^2 / (4 \pi)^2$ and the standard group theoretic symbols are specified 
in appendix \ref{app:QCD-like}.  In principle we could quote ${\cal O}(\alsp{2})$ but refrain from doing
so since we believe that there is no further insight to be gained from it. 
The $\MS$-metric is obtained by using \eqref{eq:ZamDef} 
and  identifying $\lDMS{g} = 16 Z_0$ in [Eq 4.18] \cite{Zoller:2016iam}
($\lDMS{g}  =   \rnDI{g}{g}{1} \eps^{-1}+ {\cal O}(\eps^{-2}) $)
\begin{eqnarray}
\label{eq:zamMS}
\zamMS{g}{g} &=& - \left( \frac{d}{d \ln \mu} - 2\eps \right) \lDMS{g} =   2  \pas ( \als \rnDI{g}{g}{1} )  \nonumber \\[0.1cm] 
&=&
\frac{n_g}{2 \pi^2}\left(1 + 2 \als \left( \frac{17}{2} C_A - \frac{10}{3} N_F T_F \right)  \right)     + {\cal O}(\alsp{2})  \;.
\end{eqnarray}
A few remarks are in order. Firstly, the LO expression is the same in both schemes 
and positive in accordance with  positivity in CFTs. The ${\cal O}(\als)$ coefficient differs 
but   in the absence of the knowledge of the higher terms  
no   firm conclusions can be drawn on  positivity. Nevertheless it is instructive  
to see for what number of flavours the sign of the second term changes. 
If we fix $N_c = 3$ then the critical number is $N_F^c|_{\MOM} \simeq 15.6$ and $N_F^c|_{\MS} \simeq 5.1$ in the \Nscheme- and $\MS$-scheme respectively. This indicates that the \Nscheme-scheme  is more likely to be positive than 
the $\MS$-scheme.
 In fact $N_F^c|_{\MOM} \simeq 15.6$ is very close to a 
CBZ-FP where the critical coupling is very small and  positivity can be expected to hold 
for the first few coefficients of $\zamS{g}{g}$. The difference between the \Nscheme- and $\MS$-metric at 
${\cal O}(\als)$ is due to the ${\cal O}(\als)/\eps^2$-term in the bare term.
Hence the single logarithm $\eps \ln(p^2)$, relevant to the definition of the metric,  
needs to be complemented with an additional 
 ${\cal O}(\eps)$-term  which cannot be deduced 
without further computation in order to obtain a finite result.
Yet since the ${\cal O}(\als)/\eps^2$-term equals 
$- \be_0{\cal O}(\alsp{0})/\eps$-term, the difference between the two metrics  has to be proportional 
to $\be_0$ which is easily verified
\begin{equation}
\zamS{g}{g} - \zamMS{g}{g} = \frac{n_g}{2 \pi^2}    \,  2 \be_0 \als   + {\cal O}(\alsp{2}) \;.
\end{equation}
Note, the $\be_0$-coefficient is consistent with the generic formula for a scheme change \eqref{eq:Rchange}.

\subsection{Caswell-Banks-Zaks Fixed Point} 
\label{sec:BZ}

The  CBZ-FP \cite{C74,BZ81} is a perturbative IR-FP which is analytically tractable and 
therefore often serves to illustrate conformal window studies explicitly. 
The  CBZ-FP in QCD-like theories (cf. appendix \ref{app:QCD-like} for the conventions)
is found by tuning $N_c$ and $N_f$ in some quark representation such that 
$\be(\alsIR) = 0$ with $\be$ approximated by some low order in perturbation theory and crucially 
$\alsIR$ being small.  This amounts to keeping the parameter 
$\kCBZ = -\frac{3}{2}\frac{\be_0}{N_c}  \ll 1$ small and introducing the following power counting 
$\als \sim {\cal O}(\kCBZ)$ and $\beta \sim {\cal O}(\kCBZ^2)$.

 Since $\Delta \barb$ is determined from the $2$-point function we may 
use the recent NNLO computation of the $\vev{G^2 G^2}$-correlator \cite{Zoller:2016iam} ($\TEMTO = \be /2 [G^2]$ 
in QCD-like theories) to obtain 
 $\Delta \barb$  and $\be_a$ to NNLO which is ${\cal O}(\kCBZ^4)$.
Concretely  $\Delta \barb$ is obtained from  \eqref{eq:barb2} 
\begin{eqnarray}
\label{eq:DelBarbExplicit}
\Delta \barb =  \frac{1}{8}  \int^\infty_{-\infty}  \zamMS{g}{g} \left( \frac{\be}{2} \right)^2   d\ln \mu'  =
  \frac{1}{32}  \int_0^{\alsIR}    \partial_u \left( \frac{\be}{u}   \right)u   
 \rnDI{g}{g}{1}(u) du \;,
 \end{eqnarray}
 where to deduce the second equality,  \eqref{eq:zamMS} and integration by parts were used.
 The first pole residue $ \rnDI{g}{g}{1}$, known 
from \cite{Zoller:2012qv}, is quoted in    \cite{PZ16} [section 3.4.2.]  in 
 the notation used here. 
 Using the formula above we get
\begin{eqnarray}
\label{eq:DelbarbCW}
\Delta \barb &\;=\;& \frac{- \be_1 \rnDI{g}{g}{1,0}} {64} (\alsIR)^2  \nonumber \\[0.1cm]
                      &\;-\;&          \frac{1}{96}( 2 \be_2  \rnDI{g}{g}{1,0} + \be_1  \rnDI{g}{g}{1,1}  )(\alsIR)^3   \nonumber \\[0.1cm]
                     &\;-\;&  \frac{1}{64}\left(\frac{3}{2} \be_3   \rnDI{g}{g}{1,0}  + \be_2  
                     \rnDI{g}{g}{1,1} +\frac{1}{2} \be_1   \rnDI{g}{g}{1,2} \right)(\alsIR)^4 + {\cal O}(\alsp{5}) \;.
\end{eqnarray}
Solving $\be(\als^{\IR})= 0$ up to the fourth order gives
\begin{equation}
\alsIR = -\frac{\be_0}{\be_1}\left(1+ \frac{\be_0 \be_2}{\be_1^2}+ \be_0^2 \frac{(2 \be_2^2- \be_1 \be_3)}{\be_1^4} \right)+ O(\be_0^4)  \;.
\end{equation}
Inserting this expression into \eqref{eq:DelbarbCW} and using \eqref{eq:bkappa}
the final result of this section reads
\begin{eqnarray}
\label{eq:k5}
\Delta \barb = \frac{1}{7200 \pi^2} N_c^2 \kCBZ^2 \left(1+ 2\left(\frac{7}{25}\right)^2 \kCBZ + \frac{53 \cdot 4231}{3^3 \cdot 25^4} \kCBZ^2\right)+O(\kCBZ^5) \;.
\end{eqnarray}
Note that LO and NLO expression agrees with reference \cite{JO90}. The $O(\kCBZ^4)$ term is new and 
it is observed that  the factor of $\zeta_3$ has dropped from the final expression.   
With the knowledge of the four loop expression $ \rnDI{g}{g}{1,3}$ one could easily extend this expression
 to $O(\kCBZ^5)$ by using the evaluation of the $\be$-function to five loops \cite{5loops}.
It is noted that since $\kCBZ = -3/2 \be_0/N_c >0$ in the conformal window the above expression is 
manifestly positive in accordance with \eqref{eq:mom4D}.
Effectively \eqref{eq:k5} corresponds to Euler flow difference $\Delta \be_a/2$ since it can be shown 
that in QCD-like theories 
$ \Delta \be_a =2 \Delta  \barb   + {\cal O}(\kCBZ^6) $  
\cite{PZprep17}.\footnote{It is presumably possible 
to obtain the Zamolodchikov-metric for the $\be_a$-flow, 
$ \chi^{g}_{gg} \sim G_{gg}$ (notation as in \cite{JO90} and \cite{JO13} on the LHS and RHS respectively) 
in QCD-like theories from eq.2.20 in \cite{JO13}. For a one coupling theory the antisymmetric $S_{gg} =0$,  
$\chi^\MS_{gg} =  -\chi^a_{gg} \sim  {\cal A}_{gg}  $ is known to NNLO and the knowledge of 
$\chi_{ggg}^b \sim B_{ggg}$ to NLO seems sufficient to get $\chi^{g}_{gg}$ at NNLO.}

\subsection{The \Rsscheme-Scheme in QCD-like Theories and the 
Renormalisation of $G^2$}
\label{sec:R2QCD}

It is instructive to consider the case of a QCD-like theory to understand what happens
in this \Rsscheme-scheme. From the work of Hathrell \cite{H81}, related to QED but sufficient for 
our purposes, the relevant part of the TEMT reads
\begin{equation}
\label{eqTEMTH}
\vev{\TEMT{\rho}}  = \frac{1}{4} (d-4) \vev{G^2} - (d-4) b_0  H^2+ \dots \;,
\end{equation}
in terms of bare quantities. The relation of the latter to the renormalised finite 
quantities is as follows
\begin{equation}
\label{eq:G2MS}
\frac{1}{4}(d-4) \vev{G^2}   =  \frac{\hat{\be}}{2}\vev{[G^2]}^{\MS}   + (d-4) \mu^{d-4} ( \LbMS - \frac{\be^{\MS}_b}{d-4}) H^2 + \dots \;, 
\end{equation}
where $b_0 =  \mu^{d-4}( b^\MS + \LbMS) $   and  the $\MS$-scheme dependence has been labelled.
In both equations the dots stand for terms which are not essential for our discussion.
Note that when $\vev{\TEMT{\rho}} $ is expressed in terms of renormalised 
quantities the $L_b$-term cancels  and the $(d-4) b H^2$ vanishes in the $\eps \to 0$ limit and 
$\vev{\TEMT{\rho}} = \be/2 \vev{[G^2]}^{\MS}  - \bb^\MS H^2$ + \dots.

Thus the question is what happens to this picture in the the \Rsscheme-scheme.
Taking the definition into account \eqref{eq:R2-scheme} we see that the equations 
above change to 
\begin{equation}
\frac{1}{4} (d-4) \vev{G^2} = \frac{ \hat{\be}}{2} \vev{[G^2]}^\Rs   +      \dots \;, \quad  
\end{equation}
with $b_0 =  \mu^{d-4} b^\Rs$.
When inserted in \eqref{eqTEMTH} this gives the same scheme-independent 
VEV of the TEMT  
\begin{equation}
\label{eq:fund}
\vev{\TEMT{\rho}} =  \frac{\be}{2} \vev{[G^2]}^\MS -  \bb^\MS H^2 + \dots   =  \frac{\be}{2} \vev{[G^2]}^\Rs + \dots \;, 
\end{equation} 
when 
expressed in terms of renormalised quantities in the  $\eps \to 0$ limit. 
The above reasoning can be restated as  $\hat{\be} ( \vev{[G^2]}^{\MS} -    \vev{[G^2]}^{\Rs} ) =  2 (\be_b^\MS - \be_b^{\Rs}) H^2 = 2  \be_b^\MS  H^2 $ valid up to terms previously denoted by dots.

\subsection{$\Delta \barb$ in Free Field Theory} 
\label{sec:free}

Free field theory flows are instructive and relevant since they describe the transition 
from an asymptotically free theory to the chiral broken phase of free massless goldstone 
bosons \cite{Cardy:1988cwa}. 
A higher derivative massive free field theory computation is deferred to appendix \ref{app:higher}.
 Concretely we think of a massive free field of spin $s$ consisting of $(2s+1)$ degrees of 
 freedom in the UV which decouple in the IR. Within this setup \eqref{eq:mom4D}, or the adaption 
\begin{equation}
\label{eq:calculator}
\Delta \barb = \frac{1}{8} \hat{P_2}  \Big|_{p=0}
\int d^4x \, e^{i x \cdot p}  \vev{ \TEMTO(x) \TEMTO(0)}_c
  \;,
\end{equation}
with $\hat{P}_2$ defined in \eqref{eq:P2}, 
 can be considered 
as an efficient $\Box R$-anomaly calculator provided (cf. section  \ref{sec:summary}) that the integral is 
convergent in the IR and the UV. For 
this to be the case  conformality ought to be broken  by soft terms  only.
This is the case for  the free massive
conformally coupled scalar and  fermion for which the operator-part of the TEMT are $\TEMTO = m^2 \phi^2$ and   $\TEMTO = m \bar q q $ (Dirac fermion)  respectively.

Using the formula \eqref{eq:calculator}  we get 
\begin{alignat}{3}
\label{eq:res-free}
& \Delta \barb_{(0,0)} &\;=\;&  \phantom{-} \frac{1}{8} m^4   {\cal B}_0''(0,m^2) &\;=\;&   1 \, [\text{unit}] \;,  \nonumber \\[0.1cm]
& \Delta \barb_{(\frac{1}{2},0)\oplus(0,\frac{1}{2}) }  &\;=\;&   -\frac{1}{8} m^2  ( 2 m^2 {\cal B}_0''(0,m^2)  +  {\cal B}_0'(0,m^2)  ) &\;=\;&   6 \, [\text{unit}] \;,
\end{alignat}
where $[\text{unit}]$ is a normalisation factor 
\begin{equation}
[\text{unit}]  = \frac{1}{3840 \pi^2}     \;,
\end{equation}
($2880 = 3/4 \cdot 3840$ converting to the conventions of  \cite{birrell1982quantum})   and  
\begin{equation}
  {\cal B}_0(p^2,m^2) =  \frac{ \Gamma(\eps)}{(4 \pi)^{2}}   \int_0^1 dx ( m^2 + x (1-x) p^2)^{-\eps}  \;,
\end{equation}
is  the bubble-integral for equal mass scalars
 with  primes denoting derivatives with respect to $p^2$ and $\Gamma$ is the Euler function.  It is readily seen that \eqref{eq:res-free} agrees with 
the results in the literature  \cite{birrell1982quantum} (cf. table 1 of chapter 6.3)  by taking into account the  conversion $c|_{\!\!\!\mbox{\cite{birrell1982quantum}}} = 4/3 \Delta 
\barb$ and the factor two for Dirac versus Weyl fermions. 
The convergence of the integral  presumably is in $1$-to-$1$ correspondence with 
scheme-independence of direct computation using a regularisation method to derive 
\eqref{eq:VEVTEMT}.
 For example the $\zeta$- \cite{Christensen:1978md} and dimensional-regularisation \cite{BC77} yield the 
 same result. This contrasts the case of the free non-conformally coupled scalar and 
the vector particle for which those methods yield different results. 
This is reflected here in that the formula 
\eqref{eq:calculator} is IR and UV divergent for the non-conformally coupled scalar $\TEMTO = -
\frac{1}{2} \Box 
\phi^2 + m^2 \phi^2$ and the vector particle. 
This issue  clearly deserves further study in view of the remarks at the beginning of this section.
An interesting aspect is that the scalar to Dirac fermion ration is $6$ for the $\Delta \barb $ but $11$ for $\Delta \be_a$ 
and might therefore give rise to tighter bounds.

\section{Summary and Outlook}
\label{sec:conclusions}

Amongst the Weyl-anomaly contributions  \eqref{eq:VEVTEMT} 
 the $\barb \Box R$-term has received considerably less
 attention as compared to the Weyl and the Euler term, presumably because 
 it is ambiguous $\barb \to \barb -  \frac{1}{8} \omega_0$ under 
 ${\cal L }\to {\cal L}   + (\omega_0/72 )R^2$ \eqref{eq:amb}.
Our starting point was the observation that whereas such an ambiguity is present 
in each theory it disappears in 
the  flow, $\Delta \barb \equiv \barb^\UV - \barb^\IR$, since the IR 
and UV ambiguity are identical.  
On the technical level the crucial ingredient is the UV-finiteness property of the $\vev{\TEMTO\TEMTO}$-correlator, 
discussed in our previous work \cite{PZ16}, allowing us to identify $\Delta \barb$ with a 
bare and therefore RG-scale invariant correlator \eqref{eq:barb0}. 
The quantity $\Delta \barb$ describes the global flow 
properties, cf. section \ref{sec:global}, which include scheme-independence and positivity $\Delta \barb >  0$ 
which are most clearly seen from the spectral representation \eqref{eq:barb1} as previously 
 observed \cite{A99}. 
 The integral representation of $\Delta \barb$ follows from an anomalous Ward  identity \eqref{eq:AWI}
 \begin{eqnarray}
 \label{eq:final0}
 \Delta \barb  &=&  \frac{1}{8}  \int^\infty_{-\infty} \left( \zamR{A}{B}{\RR} \be^A \be^B \right) ( \mu')  d\ln \mu'   
 \nonumber \\[0.1cm]
 & = &
 \frac{1}{8}  \int^\infty_{-\infty}  \frac{d}{d \ln \mu'}   \cwiR{\theta}{\theta}{\RR}(p,\mu')    d\ln \mu'   \; .
 \end{eqnarray}
 The integrand being a total  derivative implies   
flow-independence of $\Delta \barb$  which is one of the main results of this work.
 The quantity  $\zamR{A}{B}{\RR} $ is the 4D analogue of the Zamolodchikov-metric and 
 independence with respect to the  $\vev{O_A O_B}^\RR$-scheme is ensured 
 by  the local  quantum action principle cf. section \ref{sec:schemeD}.

 

The key point in discussing the local flow properties (cf. section \ref{sec:local}) 
is the discussion of  scheme-dependences since flows, in general, are known 
to be scheme-dependent outside fixed points. 
The definition of the Zamolodchikov-metric $\zamR{A}{B}{\RT}$ ($2$-form) 
in the \Nscheme-scheme  \eqref{eq:chiABOS} is considerably simpler than the 
generic Lie derivative   definition \eqref{eq:zamR}.  
For the former
positivity $\zamR{A}{B}{\MOM}(\mu) \geq 0$ 
is shown to hold non-perturbatively using a spectral representation.
This suffices  to define a quantity  ($\dot{\phantom{x}} = - \frac{d}{d \ln \mu}$)
\begin{equation}
\label{eq:final1}
 \barb^\MOM_{\RB}(\mu) 
=    \barb^\UV_{\RB} -  \frac{1}{8} \int^\infty_{\ln \mu/\mu_0}  \left(\zamS{A}{B} \be^A \be^B \right) (\mu') d\ln \mu' \;,  \quad  
\dot{ \barb}^\MOM_{\RB} < 0 \;,
\end{equation}
which is monotonically decreasing along the flow \eqref{eq:mono} where 
  $\RB$ is the scheme-prescription of the   $b_0 R^2$-term  \eqref{eq:b-coupling}.
Moreover the  UV-finiteness  \cite{PZ16} allows us to define a scheme, 
referred to as the \Rsscheme-scheme,  for which 
the $R^2$-anomaly vanishes along the entire flow $\be_b^{\rsscheme} = 0$. 
In these particular schemes,  $   \barb^\MOM_\RB(\mu)$ obeys  a gradient 
flow type equation (\ref{eq:gfpp},\ref{eq:GF}) which in the notation here reads 
\begin{equation}
\label{eq:gfpp2}
\dot{ \barb}^\MOM_{\rsscheme}(\mu)  
 =  - \frac{1}{8}\zamS{A}{B}  \be^A \be^B< 0 \;.
\end{equation}
Furthermore in section \ref{sec:BZ} we extend  $\Delta \barb$ for Caswell-Banks-Zaks fixed point to NNLO using  a recent computation of the $\vev{G^2 G^2}$-correlator. This corresponds to fourth order in the Caswell-Banks-Zaks 
coupling and constitutes also an extension of the Euler flow $\Delta \be_a$ ($a$-theorem) to the same order 
since $ \Delta \be_a =2 \Delta  \barb  $ up to  the sixth order  \cite{PZprep17}.

It is noteworthy that, due to 
topological protection, $\be_a$ is well-defined at both the UV- and IR-CFT. 
 As discussed above such a term is also irrelevant 
for  $\Delta \barb$  but requires an adaptation of the moment formula \eqref{eq:mom4D} 
to  \eqref{eq:barb0}.\footnote{ 
One may distinguish a total of four scheme choices: 
the dynamical couplings $g^Q$, the $b$-coupling ($\RB$-scheme), the choice of the 
$2$-point function for the dynamical operators ($\RT$-scheme) and 
$\omega_0 R^2$-term \eqref{eq:amb}. 
Other than in section \ref{sec:schemeD} the scheme of the dynamical couplings have not been considered. The $\RB$-scheme and the $\omega_0 R^2$-term are related in that 
$b_0 = \mu^{(d-4)}(b^\RB(\mu) + L_b^\RB(\mu) + \omega_0)$ where $\omega_0$ is 
$\mu$-independent cf. appendix \ref{app:2-ways} for further remarks.}

  Generally the $\Delta \barb$-integral representations  (\ref{eq:barb0})-(\ref{eq:barb2}) are correct when 
 conformality is  broken by soft terms only, e.g. $\TEMTO = m^2 \phi^2$ and $\TEMTO = m \bar q q$, 
 in which case the integrals  converge in the IR and UV and \eqref{eq:barb0} 
can be regarded as a $\Box R$-anomaly calculator. 
 UV-convergence is ensured for asymptotically safe and asymptotically free theories \cite{PZ16}. 
 Free field theories are a class on their own, coherent with our finding that convergent correlation functions
 diverge at fixed order in perturbation theory.  
Since propagators of massive fields $\Phi^{(s)}$ of spin $s$ 
contain terms scaling like  $(k^2)^{s-1}$, the representation in  \eqref{eq:barb0}  
diverges in the UV  for conformally coupled fields of spin $1$ and higher.\footnote{This seems linked to the scheme-dependence found for direct evaluation 
 of the  spin $1$ term via \eqref{eq:VEVTEMT} cf. \cite{birrell1982quantum}.}
UV-convergent cases include  the previously quoted free spin $0$ (conformally coupled) and spin $1/2$  particles 
for which we find results (cf. section \ref{sec:free}) in accordance with direct $\Box R$-computations  
\cite{birrell1982quantum}. Non-conformal couplings of the type 
$\TEMTO = - \frac{1}{2} \Box  \phi^2 + m \phi^2$ worsen the situation and already lead to UV-divergences 
 in  \eqref{eq:barb0} for spin $0$ fields. 
 IR-divergences occur for non-conformally coupled spin $0$ fields 
 $\Delta \barb \sim \ln (m_\phi)$ (C.f. the discussion in section \ref{sec:chiral}).   
 
 The problems of a free spin $1$ particle  might be cured by using a  gauge invariant formulation, 
 e.g. providing mass  to the spin $1$ field via a Higgs-mechanism as mentioned 
 elesewhere \cite{Cappelli2}. 
 The non-conformally coupled scalar is relevant since  it is associated with the goldstone boson of a spontaneously broken chiral symmetry. 
The IR-divergence  does not appear to resum to  a finite expression cf. section \ref{sec:chiral}.
 Since chiral symmetry and conformal symmetry are regarded as excluding each other,  
 removing  the  $\Box \pi^2$-term, with $\pi$ denoting the goldstone bosons,  by the usual improvement   \cite{CCJ70} seems prohibited. 
 If the prescription $\TEMTO \to - \frac{1}{2}\Box \pi^2$ is correct then $\Delta \be_a$, the flow  of  the Euler 
 term,  still 
seems  well-defined since  its topological nature 
permits to bypass the problem in an efficient manner cf. footnote \ref{foot:practical}.   
What happens for the flow of $\Box R$ is less clear. It might either indicate that the flow 
$\Delta \barb$ diverges or that the formulae need to be amended.
It is possible that this situation  may change should there exist a phase 
where scale symmetry is spontaneously broken (Goldstone-Nambu realisation) 
and the pion degrees of freedom are supplemented by a dilaton in which case improvement   
might be possible. 
Clearly the question of IR-divergencies of the chirally broken phase  deserves further study.\footnote{So does 
a systematic study of  dimensionful  couplings, e.g. \cite{JO13} for local RG-formulations, beyond the remarks in section \ref{sec:convergence}.}
The resolution for   the $\Box R$-flow has the potential to render it  more predictive 
for theories with broken chiral symmetry, e.g.  a bound on the conformal window which differs from 
the one of the $a$-theorem.

\appendix
\numberwithin{equation}{section}

\subsection*{Acknowledgements}

We are grateful to  Roberto Auzzi, Luigi Del Debbio, Martin Evans, Gino Isidori, Tony Kennedy, Zohar Komargodski, Anatoly Konechny, Donal O'Connell, Hugh Osborn, 
Toni Pich, Adam Schwimmer, Graham Shore, Andreas Stergiou and Lewis Tunstall for useful discussions and 
Saad Nabeebaccus \& Ben Pullin for thorough proofreading of the manuscript. 
VP would like to thank CP3-Origins for hospitality during final  
stages of this work.

\section{Derivations of $\Delta \barb  \sim
\int d^4 x\, x^4   \vev{ \TEMTO(x) \TEMTO(0)}_c $ }
\label{app:derivations}

In this appendix we derive the fourth moment formula for $\Delta \barb$ \eqref{eq:mom4D}
using anomalous WIs, the (Weyl) anomaly matching procedure by Komargodski and Schwimmer \cite{KS11,K11}
and indirectly by veryfing \eqref{eq:barbdmu} for QCD-like theories using result 
by Hathrell  \cite{H81} on the renormalisation of the field strength tensor in curved space
in sections \ref{app:AWI}, \ref{app:athm} and \ref{app:QCD-like} respectively. 
We stress that the derivations of in section \ref{app:AWI} and \ref{app:athm} 
are general and do not rely on the specific interplay of $\sigma$ and $b$ in QCD-like theories.

\subsection{The fourth Moment and $\Delta \barb$ from an Anomalous Ward Identity}
\label{app:AWI}

Anomalous WIs can be obtained by applying operator combinations 
of the form ${\cal D}_{\pm }(x,\mu)  \equiv - \left( \de_{s(x)} \pm \be^A \de_{A(x)}(\mu) \right)$  to the partition function. 
A single application gives 
\begin{equation}
\label{eq:up}
{\cal D}_{- }(x,\mu) \, \ln Z = \sqrt{\gm} ( \vev{ \TEMT{\rho}(x) }^\RR  - \be^A \vev{ [O_A(x)]}^\RR )= 
4  \barb^\RR   \,   \sqrt{\gm}\Box H   + \dots \;,
\end{equation}
where the dots stand for  terms which cancel from the final expression.  The quantity $\gm$ denotes the determinant of the 
metric $\gm_{\al \be}$.
Note, the $\mu$-dependence of $\barb$ is balanced on the LHS by the second term.
The WIs are anomalous in the sense that they display the Weyl 
anomaly 
on the RHS of \eqref{eq:up}. Applying a second ${\cal D}$-operator leads to 
\begin{eqnarray}
\label{eq:AWI}
 {\cal D}_{+ }(0,\mu)  {\cal D}_{- }(x,\mu)  \, \ln Z|_{\gm_{\al  \be} \to \de_{\al \de}}   &=& 
     \left( \vev{\TEMT{\rho}(x) \TEMT{\la}(0)}_c^\RR  - \be^A \be^B \vev{ [O_A(x)] [O_B(0)] }_c^\RR \right)  +
 \nonumber  \\[0.1cm]
& \phantom{+}&      \left( 2 \vev{ \TEMT{\rho}(x) }^\RR  -   \be^ B (\partial_B     \be^A) \vev{ [O_A(x)]}^\RR \right) 
\de(x)    \nonumber  \\[0.1cm] 
&  =& - 8  \barb^{\RR} \,   \Box^2 \de(x)     \;,
\end{eqnarray}
where  the vanishing of the  commutator,  $[\de_{s(x)} , \be^A \de_{A(0)}]=0$, was used. The anomalous WI \eqref{eq:AWI} corresponds to eq.~5.21  in \cite{S16} (in Minkowski space). 
With regard to the notation  \cite{O91,S16},  the  identification 
 $4 \barb^{\RR}(\mu) \equiv 4( \sigma^\RR(\mu) - b^\RR(\mu) )  = \tilde{d}(\mu) \equiv d + \frac{1}{2} \be^Q U_Q(\mu)$ 
and $4 \barb^\IR = d $, gives a consistent picture.
Note that by combining different anomalous WIs some Weyl consistency conditions arise \cite{S16}.  This is of little surprise since the commutator above encodes the essence
of the Weyl consistency relations.

Applying $\int d^4 x \, x^4$ to  \eqref{eq:AWI}  and differentiating with respect to the scale  $\frac{d}{d \ln \mu} $
  one obtains 
\begin{equation}
\label{eq:gf}
\frac{d}{d \ln \mu}  \barb^\RR
= \frac{1}{8}  \zamR{A}{B}{\RR} \be^A \be^B   \;,
\end{equation}
upon using \eqref{eq:ZamDef}, \eqref{eq:Zamo} and $\TEMTO = \be^A [O_A]$.
Above we have directly assumed the $\eps \to 0$ limit 
and crucially used the fact that $\frac{d}{d \ln \mu} \momDD{s}^{(2),\RR}(p,\mu) =0  $, 
the renormalised counterpart of the $\vev{\TEMT{\rho}(x) \TEMT{\la}(0)}_c^\RR$-correlation function, is scale independent. This is the case because the counterterm 
$b_0$ in \eqref{eq:Lgrav}   is scale independent.  
Combining eqs.~\eqref{eq:ZamDef} and \eqref{eq:gf} one obtains eq.~\eqref{eq:Dbarbw0}, 
$\Delta \barb    =  \frac{1}{8} ( \momDD{\theta}^{(2)}(0)- \momDD{\theta}^{(2)}(\infty) ) $, with   more detail 
shown in  section \ref{sec:global}, which is equivalent to \eqref{eq:mom4D} and completes the task of this appendix.

\subsection{The fourth Moment and $\Delta \barb$ \`a la Komargodski and Schwimmer}
\label{app:athm}

The fourth moment formula for $\Delta \barb$ \eqref{eq:mom4D} is derived here 
in close analogy to the second moment formula for $\be_c^{2D}$ in \cite{K11} 
building  on the anomaly matching  procedure in \cite{KS11}. 
The derivation proceeds by matching the  term $ \barb^\IR$ in the IR effective action 
\begin{equation}
\label{eq:IReff}
\ln \Zpart = - \barb^\IR \int d^4 x \sqrt{g} H^2  + \dots \;,
\end{equation}
with the path integral expression.  Above  
the dots stand for non-local and Weyl-invariant contributions.   
The local part of \eqref{eq:IReff} is dictated by the IR trace anomaly \eqref{eq:VEVTEMT}.
The correctness of \eqref{eq:IReff} follows from a Weyl-variation
 $\gm_{\mu \nu} \to  e^{-2 s(x)} \gm_{\mu \nu}$ for which 
$\vev{\TEMT{\rho}} = (-\de_{s(x)}) \ln \Zpart$ and $ (-\de_{s(x)}) H^2 = 4 \Box H$.
In what follows  it is convenient to assume a conformally flat background  $\gm_{\mu \nu}= e^{-2 s(x)} \delta_{\mu \nu}$ for which 
\begin{equation}
\label{eq:IReffCF}
\ln \Zpart = - 4 \barb^\IR \int d^4 x (\Box s)^2 + {\cal O}(s^3) \;.
\end{equation}
One might wonder whether the presence of $W^2$ and $E_4$ would interfere in this picture. 
This is not the case since for conformally flat background $W^2$ vanishes and 
$E_4$ does not contain a quadratic term in $s(x)$. In passing we remark that this fact is at the heart of the difficulty of establishing the 4D $a$-theorem ($ \Delta \ba \geq 0$).

On the other hand $\ln \Zpart$ written as the Euclidean path integral over dynamical fields $\phi_i$ reads
\begin{equation}
\label{eq:logZs}
\Zpart =  \left( \int \mathcal{D}\phi_i e^{-S_{dyn}(\phi_i,\gm_{\mu \nu}) + b_0 \int d^4 x \sqrt{g} H^2} \right) 
=
\left( \int \mathcal{D}\phi_i  e^{-S_{dyn}(\phi_i,s) + 4 b_0\int d^4 x (\Box s)^2 + {\cal O}(s^3)} \right) \;,
\end{equation}
where the conformally flat metric was assumed in the second equality and 
$b_0$ is the bare gravitational counterterm with conventions specified in \eqref{eq:Lgrav}. Note that these conventions imply a somewhat unfortunate sign 
of the initial condition $b_0 = - \barb^\UV$.\footnote{\label{foot:help} 
It is instructive to  underlay this statement in the language of the QCD-like example of section 
\ref{app:QCD-like}.
Using  $d = 4$, the following lengthy chain applies
of equations   
$b_0 = b^\UV + L_b^\UV  = b^\UV = - ( \sigma^\UV - b^\UV) = - \barb^\UV$, 
when taking into account that $L_b^\UV = 0 $ and $\sigma^\UV = 0$.}
The quantity $\barb^\IR$ is found by  performing a derivative expansion of the quantum part of the path integral  in order to match the $(\Box s)^2$-term in \eqref{eq:IReffCF}. Concretely 
\begin{eqnarray}
\label{eq:exps}
& & \ln  \int \mathcal{D}\phi_i e^{-S_{dyn}(\phi_i,s)}= \nonumber \\
& &  \ln \Zpart_0 - \int d^4 x s(x) \vev{\TEMTO(x)}
 + \frac{1}{2}\int \int d^4 x d^4 y s(x)s(y) \vev{\TEMTO(x) \TEMTO(y)} + {\cal O}(s^3) \;,
\end{eqnarray} 
where here $\vev{\dots}$ refers to the  flat-space VEV.  The TEMT correlators appear in the expansion since 
$s(x)$ is the source term for the latter.
The four derivative term \eqref{eq:IReffCF}
is matched by Taylor expanding   the double integral term in \eqref{eq:exps} to fourth order 
\begin{equation}
s(y)=s(x)+ \dots + \frac{1}{4!} (x-y)^{\mu}(x-y)^{\nu}(x-y)^{\rho}(x-y)^{\sigma}\partial_{\mu}\partial_{\nu}\partial_{\rho}\partial_{\sigma}s(x)+ {\cal O}(\partial^5) \;.
\end{equation}
Using the Euclidean rotational symmetry the following replacement 
\begin{equation}
(x-y)^{\mu}(x-y)^{\nu}(x-y)^{\rho}(x-y)^{\sigma} \to \frac{1}{24} (x-y)^4( \delta^{\mu \nu} \delta^{\rho \sigma}+\delta^{\mu \rho} \delta^{\nu \sigma}+ \delta^{\mu \sigma} \delta^{\rho \nu})  \;,
\end{equation}
is valid under the  integral.
Changing the integration variable to $y= z+x$ one gets
\begin{align} \label{eq:DerExp} \notag
\frac{1}{2}\int \int d^4 x d^4 y s(x)s(y) \vev{\TEMTO(x) \TEMTO(y)}_c &= \frac{1}{2}\int d^4 x s(x)^2 \int d^4 z \vev{\TEMTO(z) \TEMTO(0)}_c + \dots \\
& \quad + \frac{1}{3 \, 2^7}\int d^4 x (\Box s(x))^2 \int d^4 z z^4 \vev{\TEMTO(z) \TEMTO(0)}_c  \;.
\end{align}
Substituting \eqref{eq:exps} in \eqref{eq:logZs} and using the derivative expansion \eqref{eq:DerExp} leads 
to 
\begin{equation}
\label{eq:4momDerived}
-\barb^\IR =  \ln \Zpart \big |_{4 \int (\Box s)^2} =  b_0 + \frac{1}{3 \, 2^9}\int d^4z z^4 \vev{\TEMTO(z) \TEMTO(0)}_c  \;,
\end{equation}
and
\begin{equation}
\label{eq:mom4D2}
\Delta \barb =  \barb^\UV -  \barb^\IR =  \frac{1}{2^9\, 3} 
\int d^4 x\, x^4   \vev{ \TEMTO(x) \TEMTO(0)}_c  \geq 0 \;,
\end{equation}
 then follows  by using the initial condition 
$b_0 = - \barb^\UV$  in the above equation.  It is important to note 
that this derivation implicitly relies on the theories being conformal in the UV and IR 
since $\bb^{\textrm{CFT}} = 0$ and so the $\barb$ and $\bb$ do not interfere in 
the Weyl anomaly \eqref{eq:VEVTEMT} when reduced to a conformally flat background.

Adding a term $\de {\cal L} \sim \omega_0 R^2$ \eqref{eq:amb}, resulting in $b_0 \to b_0 + \frac{1}{8}\omega_0$  does not affect \eqref{eq:mom4D2} since it is present in both the UV and IR term
 $ \barb^\UV = -  b_0$  of  $\barb^\IR = - b_0 + \frac{1}{8}   \cwiD{\theta} (0)$. 
Stated more simply $b_0$ is only an initial value which does not affect the difference accumulated in the flow. 
A more serious issue is the question as to  whether the fourth moment converges 
in the UV and IR which is discussed in section \ref{sec:convergence}.

\subsection{The fourth Moment and $\Delta \barb$ \`a la Hathrell in QCD-like Theories}
\label{app:QCD-like}
In this section we rederive the formula  \eqref{eq:barbdmu} in QCD-like theories  by direct use 
of the expressions for $\bb^\MS$ \& $\lDMS{\theta}$ , the local QAP and results on the renormalisation of $G^2$ in the external gravitational field.
The link between the gravity counterterms  \eqref{eq:Lgrav} 
and  $\vev{ \TEMTO \dots  \TEMTO}$-correlators 
  is given by the QAP and establishes 
  $\LbMS  =- \frac{1}{8}\lDMS{\theta}$ \eqref{eq:link} which consists in  our first step.
The relation between $b$ and $\barb$ is as follows 
\begin{equation}
\barb(\mu) =  \sigma(\mu)-b(\mu)   \;, \quad \sigma^\UV = 0
\end{equation}
where $\sigma(\mu) = \sigma(\als(\mu)) $ is a quantity related to the renormalisation of $G^2$ 
in a curved background \cite{H81}.\footnote{The  quarks and gluons that are integrated out
in an external gravitational field lead to a curvature term $\Box R$ which when divergent needs  
to be subtracted.}
In some more detail the bare $b_0$ in the 
Lagrangian \eqref{eq:Lgrav} (with $\eps \to 0$ allowed by finiteness of $L_b$  \eqref{eq:link}) is 
\begin{equation}
\label{eq:b0}
b_0 \equiv b^\UV = b(\mu) + L_b(\mu)  \;,
\end{equation}
where we remind the reader that the $\mu$-dependence arises from $\als(\mu)$.
From the explicit expression of $\lDMS{\theta}$ given in section 3.1 of \cite{PZ16}, it is observed  that  ($\eps \to 0$   implied)  
\begin{eqnarray}
\label{eq:Lbmu}
\LbMS(\mu)  &\!=\!&  - \frac{1}{32}  \int_0^{\als}    \partial_u \left( \frac{\be}{u}   \right)u \left(1- \frac{u}{\als} \right)  \rnDI{g}{g}{1} (u) du  \nonumber \\[0.1cm]
 &\!=\!&  \frac{ \bebMS}{ 2\be}  - \frac{1}{32}  \int_0^{\als}    \partial_u \left( \frac{\be}{u}   \right)u  \rnDI{g}{g}{1} (u) du \nonumber \\[0.1cm]   
&\!=\!&  -\sigma^\MS  - \frac{1}{32}  \int_0^{\als}    \partial_u \left( \frac{\be}{u}   \right)u  \rnDI{g}{g}{1} (u) du     \;,
\end{eqnarray}
where in the last line the formula $\sigma  =- \be_b /( 2\be)$  \cite{H81} was used along with
  the formula for $\be_b$\footnote{From \eqref{eq:beb-g1} 
one infers that $\bb^\MS = {\cal O}( \alsp{3})$ since $\rnDI{g}{g}{1} = {\cal O}(\alsp{0})$ 
 and that the $R^2$-anomaly-term is absent for theories with $\be = - \be_0 \als$ which is the case 
 for ${\cal N}=1$ supersymmetric Yang-Mills theory.}
\begin{equation}
\label{eq:beb-g1}
\bebMS = - \left( \frac{d}{d \ln \mu} - 2\eps \right)  \LbMS =  
\frac{1}{16}  \frac{\beta(\als)}{\als} \int^{\als}_0 \partial_u \left( \frac{\beta(u)}{u} \right)   u^2 \rnDI{g}{g}{1}(u) du  \;.
\end{equation}
Taking the IR limit ($\als \to \als^\IR$) in  \eqref{eq:Lbmu} we get
\begin{equation}
\LbMS(\alsIR) = -\sigma^{\IR,\MS} - \frac{1}{32}  \int_0^{\als^\IR}    \partial_u \left( \frac{\be}{u}   \right)u  \rnDI{g}{g}{1} (u) du   \;.
\end{equation}
Further using $L_b(\alsIR) =   b^\UV - b^\IR$  \eqref{eq:b0} and taking into account $\sigma^{\textrm{UV}} =0$ one arrives at 
\begin{equation}
\label{eq:CbFormula}
\Delta \barb = \frac{1}{32}  \int_0^{\als^\IR}    \partial_u \left( \frac{\be}{u}   \right)u  \rnDI{g}{g}{1} (u) du   \;,
\end{equation}
in  agreement with \eqref{eq:DelBarbExplicit}. 
Since the latter is equivalent to the fourth moment \eqref{eq:barb0}  the task of this section  is completed. 

\section{On the Asymmetric-part to the Gradient Flow equation \eqref{eq:GF}}
\label{app:asym}

The goal of this appendix is to discuss the antisymmetric part 
$ \AS_{AB}$ in \eqref{eq:GF}. 
Clearly such a term does not affect global results since it vanishes 
when contracted by $\be^A \be^B$ in \eqref{eq:gfpp}. 

The possibility of such a term can be inferred directly from the definition 
$  \zamR{A}{B}{\RR} = - {\cal L}_\be   \lnD{A}{B}$ \eqref{eq:zamR}, related to $\GM_{AB}$ as in 
\eqref{eq:short}. It is straightforward to obtain 
\begin{equation}
\label{eq:fini}
\be^A \zamR{A}{B}{\RR}  = - \partial_B f^\RR - \be^A \ASz^\RR_{AB} \;,
\end{equation}
where 
\begin{equation}
f^\RR =  \lnD{\theta}{\theta} = \be^A \be^B  \lnD{A}{B} \;, \quad \ASz^\RR_{AB} = \partial_{[A} F^\RR_{B]} \;,
\end{equation}
with $F_B^\RR = \be^C \lnD{C}{B}$ and the square bracket denoting antisymmetrisation in the indices 
$A$ and $B$ as usual.  Now 
\begin{eqnarray}
\partial_{A} F^\RR_{B} &=&  \partial_A  \be^C \lnD{B}{C} + \be^C \partial_A \lnD{B}{C}  \nonumber \\[0.1cm]
&=&  \ga_A^{\phantom{A}C} \lnD{B}{C} + \be^C \partial_A \lnD{B}{C} \;,
\end{eqnarray}
whose antisymmetric part is not obviously vanishing. Hence at this formal level
the vanishing of $ \AS_{AB} = \frac{1}{ 8} \ASz^\RR_{AB} =  \frac{1}{8}\partial_{[A} F^\RR_{B]}$ 
cannot be concluded and $ \AS_{AB} $ has therefore to be included in \eqref{eq:GF}. 
The antisymmetric part is the reason why equation \eqref{eq:GF} is referred to as gradient flow type  rather 
than gradient flow only. 

An interesting question is as to whether $\ASz^\RR_{AB}$ is finite or not. 
Eq.~\eqref{eq:fini} implies so since $\be^A \zamR{A}{B}{\RR}$ and $\partial_B f^\RR$ are both finite. 
The former is finite since $\zamR{A}{B}{\RR}$ is an anomalous dimension of the $\vev{O_AO_B}$-correlator 
and $\partial_B f^\RR$ is the derivative of the finite quantity $f^\RR =  \lnD{\theta}{\theta}$ \cite{PZ16}.  
In eq.~\eqref{eq:fini} an evanescent term proportional to $2 \eps \be^A L_{AB}$ was omitted 
which comes from the $d$-dimensional relation $  \zamR{A}{B}{\RR} = (2 \eps - {\cal L}_\be)   \lnD{A}{B}$  e.g.  \eqref{eq:ZamDef}. Such a term can though safely be neglected since 
it is finite even in the free field theory limit. 
On a final note, the relation to Osborn's formalism is that $\partial_{[A} F_{B]}  \sim \partial_{[A} U_{B]}$ in the notation used in the Weyl consistency 
paper  \cite{O91} and the formula 
$F_B^\RR = \be^C \lnD{C}{B}$ resembles the one given in eq.~2.17 in 
\cite{Behr:2013vta} in the 2D case. 

\section{Different Ways of handling the Gravity Counterterms}
\label{app:2-ways}

The gravity counterterms ${\cal L}_{\textrm{gravity}} = -( a_0 E_4 +   b_0  H^2 + c_0 W^2 )$ 
\eqref{eq:Lgrav} are not always treated uniformly  in the literature.
We first describe the two different ways and then show that they give rise to equivalent 
RG equation for the VEV of the TEMT.\footnote{
This is our interpretation on the topic which emerged from  illuminating
exchange with Hugh Osborn.}
\begin{enumerate}
\item The authors of references \cite{BC80,H81,F83} and ourselves (cf. section \ref{sec:R2}) impose $\frac{d}{d \ln \mu} v_0 = 0$ 
for $v = a,b,c$ therefore treating the coefficients of the gravity terms like regular couplings. 
This leads to $\frac{d}{d \ln \mu} \vev{\TEMT{\rho}} = 0$ for 
the generally accepted definition of $  \vev{\TEMT{\rho}}$ \eqref{eq:VEVTEMT}.
\item Jack and Osborn decide not to treat $v_0$ as couplings but as pure counterterms (choosing 
the $\MS$-scheme in particular), which translates in our notation to $v_0 = \mu^{(d-4)} L_v$. This then obviously 
leads to $\frac{d}{d \ln \mu} \vev{\TEMT{\rho}} \neq  0$. 
\end{enumerate}
Hence one might wonder whether these two ways of dealing with the gravity counterterms  are reconcilable. 
In fact, as Jack and Osborn remark, below eq.~2.8 \cite{JO90}, these two ways are equivalent.
Let us see how this works, assuming  that the $a_0$ and the $c_0$ terms are not present which 
simplifies the presentation.
 In our way (item 1) the RG equation for the VEV of the TEMT 
is homogeneous and reads
\begin{equation}
\label{eq:homo}
\frac{d}{d \ln \mu} \vev{\TEMT{\rho}} =  
\left( \frac{\partial}{\partial \ln \mu}  + \be^A \partial_A + \be_b \partial_b  \right)
\vev{\TEMT{\rho}}\Big|_{ b_0 = \mu^{(d-4)}(b  + L_b)}  = 0 \;.
\end{equation}
If  treated \`a la Jack and Osborn (item 2) 
the RG equation is  inhomogeneous  
\begin{equation}
\label{eq:inhomo}
\frac{d}{d \ln \mu} \vev{\TEMT{\rho}} =  
\left( \frac{\partial}{\partial \ln \mu}  + \be^A \partial_A  \right)
\vev{\TEMT{\rho}}\Big|_{ b_0 = \mu^{(d-4)} L_b}  =   4 \be_b  \Box H  + \dots\;,
\end{equation}
where $\be_b =   \frac{d}{d \ln \mu} L_b   $ was used. Note that the $\partial_{\ln \mu}$-terms vanish in mass-independent schemes as assumed in this work.
Now \eqref{eq:homo} is seen to be equivalent 
upon noticing that $\vev{\TEMT{\rho}} = 4 \barb \Box H + \dots $  and using that $\partial_b \barb = -1$. 
At last we would like to state that it is our understanding that in both formalisms one can add an arbitrary ($\mu$-independent) constant to $b_0 \to b_0 + \mu^{d-4}\frac{1}{8} \omega_0$.  
This constant term is related to the famous $\Box R$-ambiguity in the trace anomaly 
\cite{birrell1982quantum,D77,BC77,AGS05,VFGS15,Chu16} which arises in tree-level computations 
in form of scheme-ambiguities. 
Note that if $\omega_0$ was $\mu$-dependent then one would deduce different conclusions from the RG-equations.
Let us note at last that the $\mu$-independence of $\vev{\TEMT{\rho}} $ might be of importance  
for the possibility of defining the gluon condensate as the derivative of the cosmological constant term
with respect to the  renormalised coupling  $\vev{[G^2]}^\RR = - 2 \partial_{\ln g^\RR} \Lambda^{\textrm{IR}}$  \cite{GC1,GC2}.

\section{Flow-independence of a Higher Derivative Free Theory}
\label{app:higher}

In reference \cite{A01eta} the higher derivative theory, of the Lee-Wick type \cite{LeeWick}, was considered
\begin{equation}
\label{eq:Lhd}
{\cal L}_{\textrm{hd}} =  \frac{1}{2} \big( (\partial \phi )^2  + m ^2 \phi^2 + \frac{( \Box \phi)^2}{M^2} \big) \;.
\end{equation}
It was found that the $\Box R$-flow is dependent on the ratio $m/M$ and therefore 
not flow-independent  \cite{A01eta}. The ratio of masses defines different trajectories in the coupling 
space e.g.  fig.~\ref{fig:flow} for an illustration.
Below we present a  conformally coupled  extension of this model which leads to a 
flow-independent result  in accordance with our findings in section \ref{sec:global} (for dimensionless couplings). 
In summary 
\eqref{eq:Lhd} can be written in terms of two free massive fields one of them with 
 negative norm.  This is of no major concern since Lee-Wick field theories are known to
 be unitary in all examples at least at the one-loop level. The standard conformal  
  $R \phi^2$-improvement is applied to each field separately. The $\Box R$-flow is then 
  given by just  twice the value  for the free scalar field 
\eqref{eq:res-free} which is in particular mass-independent.

The solution of the \eom\;of \eqref{eq:Lhd}  shows that the $2$-point function propagates two 
degrees of freedom ($m_{1,2}^2 = (M^2/2) ( 1 \mp \sqrt{1 - 4 m^2/M^2})$)
\begin{equation}
\int d^4 x e^{i x \cdot p} \vev{\phi(x) \phi(0)} = 
\frac{M^2}{(p^2 + m_1^2)(p^2+m_2^2)} =  \frac{M^2}{m_2^2 - m_1^2} \left(  \frac{1}{p^2 + m_1^2} -  
 \frac{1}{p^2 + m_2^2} 
\right)   \;.
\end{equation} 
These two degrees of freedom can be made explicit at the Lagrangian level
by  introducing an auxiliary field $\chi'_2$ \cite{GOW}
\begin{equation}
\label{eq:Laux}
{\cal L}_{\textrm{aux}} = \frac{1}{2} \big( (\partial \phi )^2  + m ^2 \phi^2 - M^2 ( \chi'_2)^2  + 2 \chi'_2 \Box \phi \big) \;.
\end{equation}
 Upon using the \eom\;$\chi'_2 = (\Box/M^2) \phi$ of \eqref{eq:Laux},  one recovers \eqref{eq:Lhd}. 
 An even more convenient form is obtained by substituting  
 $\phi = \chi'_1 + \chi'_2$ 
\begin{eqnarray}
\label{eq:L12}
{\cal L}_{12} &\;=\;& \frac{1}{2} \big( (\partial \chi'_1) ^2 -  (\partial \chi'_2 )^2   + m ^2 (\chi'_1 + \chi'_2)^2 
- M^2  (\chi'_2)^{2} \big) 
 \nonumber \\[0.1cm]
&\;=\;& \frac{1}{2} \big((\partial \chi_1) ^2 -  (\partial \chi_2 )^2   + m_1^2 \chi_1 ^2
- m_2^2  \chi_2^2    )  
 \big) \;.
\end{eqnarray}
In the second line we have passed to   the mass eigenstates, $m^2_{1,2}$ quoted above, 
 by a hyperbolic rotation  conserving the 
 kinematic structure.  
It is apparent that $\chi_1$ and $\chi_2$ correspond to  free massive  positive and a negative normed states 
respectively.  
 The two scalar fields can be conformally coupled 
 by  the standard technique   ($\eta = \frac{1}{6}$) \cite{CCJ70} 
\begin{equation}
\label{eq:conf}
{\cal L}^{\textrm{conf}}_{12} =  \frac{1}{2} \big((\partial \chi_1) ^2 -  (\partial \chi_2 )^2   + m_1^2 \chi_1 ^2
- m_2^2  \chi_2^2      +  \eta R  (\chi_1^2 - \chi_2^2 )   \big) \;.
\end{equation}
Conformality  can be made manifest for a conformally flat metric 
$\gm_{\al \be} = e^{-2s(x)} \de_{\al \be}$ introducing the Weyl-invariant fields 
$\bar{\chi}_{1,2} = e^{-s} \chi_{1,2}$.   The function $s(x)$ conveniently act as a source for the TEMT. 
The action $S^{\textrm{conf}}_{12}[s] = \int d^4 x \sqrt{\gm}{\cal L}^{\textrm{conf}}_{12}   $  assumes the form ($\Delta\eta  \equiv (\eta - \frac{1}{6})$)
\begin{equation}
\label{eq:Sconf}
S^{\textrm{conf}}_{12}[s] = \frac{1}{2}  \int d^4 x   \big( (\partial \bar{\chi}_1) ^2 -  (\partial \bar{\chi}_2 )^2   + \bar{m}_1 ^2 
\bar{\chi}_1^2
- \bar{m}_2^2  \bar{\chi}_2^2      +    \Delta \eta \bar{R}(\bar{\chi}_1^2 - \bar{\chi}_2^2 )   \big) \;,
\end{equation} 
where $\sqrt{\gm} = e^{-4 s}$ has been absorbed into 
$\bar m_{1,2}  = e^{-s} m_{1,2}$,  $\bar{R} =6 (\Box s - (\partial s)^2)$  
 and here and below $(\partial \chi )^2 = \de^{\al \be}  \partial_{\al} \chi  \partial_{\be} \chi  $ is understood to be 
 contracted with the flat metric. 
Crucially, the action \eqref{eq:Sconf} is manifestly conformally invariant 
for $\eta = \frac{1}{6}$ up to the mass terms which break the symmetry softly. 
 The TEMT then follows from 
\begin{eqnarray}
\vev{\TEMTO(x)}  = ( - \bar{\de}_{s(x)}) |_{s= 0  }  \ln \Zpart =  m_1 ^2  \chi_1^2
- m_2^2  \chi_2^2  + {\cal O}(\Delta \eta)  \;,
\end{eqnarray}
where $\bar{\de}_{s(x)}$ indicates that $\bar{\chi}_{1,2}$ but not $\bar{m}_{1,2}$  are kept fixed.
This is the TEMT of two free massive fields for which $\Delta \barb$ is then simply twice the result 
of a free field \eqref{eq:res-free} 
\begin{equation}
\label{eq:bindep}
\Delta \barb|_{{\cal L}_{\textrm{hd}}} = 2 \Delta \barb_{(0,0)} = 2 [\text{unit}]  \;.
\end{equation}
It is interesting to note that the negative norm state gives a positive contribution to the $\Box R$-flow. 
This is intimately tied to the fact that Lee-Wick theories are unitary (at least at one-loop). 
Most importantly we find, contrary to \cite{A01eta}, that this model is independent 
of the mass ratio and therefore flow-independent. 
 
At last it might be instructive to  give the conformally coupled higher derivative version 
of the action \eqref{eq:Sconf} by performing the previous steps backwards 
\begin{equation}
\label{eq:Sconfhd}
S^{\textrm{conf}}_{\textrm{hd}}[s] = \frac{1}{2}  \int d^4 x   \left( (\partial \bar{\phi}) ^2    + \bar{m} ^2 \bar{\phi}^2 
+ \frac{( \Box \bar{\phi})^2 }{\bar{M}^2}  +  
 \Delta \eta  \bar{R}  \left(\bar{\phi}^2 \left(1 +  \frac{\Delta \eta \bar{R}}{\bar{M}^2}  \right)  - 
 \frac{2\bar{\phi}  \Box \bar{\phi}}{\bar{M}^2} \right)   \right) \;,
\end{equation} 
where  $\bar{M}^2 = \bar{m}_1^2 + \bar{m}_2^2$ was used.
The corresponding higher derivate TEMT assumes the form
\begin{eqnarray}
\label{eq:TEMThd}
\vev{\TEMTO(x)}  = ( -\bar{ \de}_{s(x)}) |_{s= 0  }  \ln \Zpart_{\textrm{hd}}  
 =
  m ^2 \phi^2 
-  \frac{(\Box \phi)^2}{M^2}    + {\cal O}(\Delta \eta) \;,
\end{eqnarray}
which one would naively expect from an improved version of \eqref{eq:Lhd}.  
 Eq.~\eqref{eq:TEMThd} differs  from the expression 
given in \cite{A01eta}. We have checked by explicit computation that \eqref{eq:TEMThd} (or \eqref{eq:Sconfhd}) 
with \eqref{eq:mom4D} give the same result as in \eqref{eq:bindep}.

\section{Convention for the QCD-like $\be$-function}
\label{app:beta}

In this work the bare $\be$-function $\hat{\be}$ of DR are defined by
\begin{equation}
\label{eq:belog}
\hat{\beta}= \frac{d \ln g}{d \ln \mu}= \frac{(d-4)}{2} + \beta  =  - \eps + \be \;.
\end{equation}
The logarithmic $\be$-function \eqref{eq:belog} is convenient for QCD and is to do 
with the unusual appearance in the Langrangian ${\cal L} = \frac{1}{4 g_0^2}  G^2$.
For multiple couplings  ${\cal L} = g^{Q}_0 O_Q $ the linear $\be$-function guarantees 
that $\be^A = \frac{d}{d \ln \mu} g^A$ transforms like a vector in coupling space.
We parameterise
\begin{equation}
\label{eq:bei}
\beta= - \beta_0 \als - \beta_1 \als^2 - \beta_2 \als^3 - \beta_3 \als^4  \dots \;, \quad \als = \frac{\al_s}{4 \pi} =  \frac{g^2}{(4 \pi)^2}
\end{equation}
where $\be_{0-3}$   in $\overline{\MS}$-scheme can be found in Ref.~ \cite{CZAKON}.
The first two coefficients, which are universal in mass-independent scheme, read 
$$\beta_0 = ( \frac{11}{3} C_A- \frac{4}{3} N_F T_F) \;, \quad  \beta_1 =  (\frac{34}{3} C_A^2- \frac{20}{3} N_c  N_F T_F - 4 C_F T_F  N_F)   \;, $$  
 where  $C_F$, $C_A$ are quadratic Casimir operators of the fundamental (quark) and adjoint (gluons) 
 representations, $N_F$ the number of quarks and $ {\textrm{tr}}[T^a T^b] = T_F \delta^{ab}$ is a 
 Lie algebra  normalisation factor  of the fundamental representation. These factors are given by
 \begin{equation}
\label{eq:SUNc}
  C_A = N_c \;, \quad  C_F = \frac{N_c^2-1}{2 N_c}  \;, \quad T_F = \frac{1}{2}  \;,
 \end{equation}
 for an $SU(N_c)$ gauge group. 
 
 \subsection{The Caswell-Banks-Zaks Fixed Point}
 
The CBZ-FP \cite{C74,BZ81} corresponds to a large $N_c,N_f$ limit with $N_f= \frac{11}{2}N_c - \kCBZ N_c$ and $\kCBZ \ll 1$. The $O(\kCBZ^4)$ calculation in sec. \ref{sec:BZ} corresponds to
\begin{eqnarray} \notag
\label{eq:bkappa}
\be_0&=& -\frac{2}{3} \kCBZ N_c \;; \quad \be_1= -(\frac{25}{2}-\frac{13}{3} \kCBZ) N_c^2 \;; \quad \be_2=-(\frac{701}{12}-\frac{53}{6} \kCBZ ) N_c^3  \;; \\
\be_3&=& (\frac{14 731}{144}+ 275 \zeta_3)N_c^4 \;;
\end{eqnarray}
where $\be_3$ was given in \cite{CZAKON}.

\bibliographystyle{utphys}
\bibliography{input4}

\providecommand{\href}[2]{#2}\begingroup\raggedright\begin{thebibliography}{10}

\bibitem{Cardy:1988tj}
J.~L. Cardy, ``{The Central Charge and Universal Combinations of Amplitudes in
  Two-dimensional Theories Away From Criticality},''
\href{http://dx.doi.org/10.1103/PhysRevLett.60.2709}{{\em Phys. Rev. Lett.}
  {\bfseries 60} (1988) 2709}.

\bibitem{Zamolodchikov:1986gt}
A.~B. Zamolodchikov, ``{Irreversibility of the Flux of the Renormalization
  Group in a 2D Field Theory},'' {\em JETP Lett.} {\bfseries 43} (1986)
  730--732.
[Pisma Zh. Eksp. Teor. Fiz.43,565(1986)].

\bibitem{Cappelli1}
A.~Cappelli, D.~Friedan, and J.~I. Latorre, ``{C theorem and spectral
  representation},''
\href{http://dx.doi.org/10.1016/0550-3213(91)90102-4}{{\em Nucl. Phys.}
  {\bfseries B352} (1991) 616--670}.

\bibitem{anomaly1}
C.~Closset, T.~T. Dumitrescu, G.~Festuccia, Z.~Komargodski, and N.~Seiberg,
  ``{Comments on Chern-Simons Contact Terms in Three Dimensions},''
  \href{http://dx.doi.org/10.1007/JHEP09(2012)091}{{\em JHEP} {\bfseries 09}
  (2012) 091},
\href{http://arxiv.org/abs/1206.5218}{{\ttfamily arXiv:1206.5218 [hep-th]}}.

\bibitem{PZ16}
V.~Prochazka and R.~Zwicky, ``{On Finiteness of 2- and 3-point Functions and
  the Renormalisation Group},''
\href{http://arxiv.org/abs/1611.01367}{{\ttfamily arXiv:1611.01367 [hep-th]}}.

\bibitem{Cappelli2}
A.~Cappelli, J.~I. Latorre, and X.~Vilasis-Cardona, ``{Renormalization group
  patterns and C theorem in more than two-dimensions},''
  \href{http://dx.doi.org/10.1016/0550-3213(92)90119-V}{{\em Nucl. Phys.}
  {\bfseries B376} (1992) 510--538},
\href{http://arxiv.org/abs/hep-th/9109041}{{\ttfamily arXiv:hep-th/9109041
  [hep-th]}}.

\bibitem{A99}
D.~Anselmi, ``{Anomalies, unitarity and quantum irreversibility},''
  \href{http://dx.doi.org/10.1006/aphy.1999.5949}{{\em Annals Phys.} {\bfseries
  276} (1999) 361--390},
\href{http://arxiv.org/abs/hep-th/9903059}{{\ttfamily arXiv:hep-th/9903059
  [hep-th]}}.

\bibitem{Zee81}
A.~Zee, ``{A Theory of Gravity Based on the {Weyl-Eddington} Action},''
\href{http://dx.doi.org/10.1016/0370-2693(82)90749-3}{{\em Phys. Lett.}
  {\bfseries B109} (1982) 183--186}.

\bibitem{birrell1982quantum}
N.~Birrell and P.~Davies, {\em Quantum Fields in Curved Space}.
\newblock Cambridge Monographs on Mathematical Physics. Cambridge University
  Press, 1982.
\newblock \url{https://books.google.co.uk/books?id=-b4LAQAAQBAJ}.

\bibitem{JO90}
I.~Jack and H.~Osborn, ``{Analogs for the $c$ Theorem for Four-dimensional
  Renormalizable Field Theories},''
\href{http://dx.doi.org/10.1016/0550-3213(90)90584-Z}{{\em Nucl.Phys.}
  {\bfseries B343} (1990) 647--688}.

\bibitem{H81}
S.~Hathrell, ``{Trace Anomalies and {QED} in Curved Space},''
\href{http://dx.doi.org/10.1016/0003-4916(82)90227-5}{{\em Annals Phys.}
  {\bfseries 142} (1982) 34}.

\bibitem{S16}
G.~M. Shore, ``{The c and a-theorems and the Local Renormalisation Group},''
\href{http://arxiv.org/abs/1601.06662}{{\ttfamily arXiv:1601.06662 [hep-th]}}.

\bibitem{KS11}
Z.~Komargodski and A.~Schwimmer, ``{On Renormalization Group Flows in Four
  Dimensions},'' \href{http://dx.doi.org/10.1007/JHEP12(2011)099}{{\em JHEP}
  {\bfseries 12} (2011) 099},
\href{http://arxiv.org/abs/1107.3987}{{\ttfamily arXiv:1107.3987 [hep-th]}}.

\bibitem{Prochazka:2015edz}
V.~Prochazka and R.~Zwicky, ``{$ \mathcal{N}=1 $ Euler anomaly flow from
  dilaton effective action},''
  \href{http://dx.doi.org/10.1007/JHEP01(2016)041}{{\em JHEP} {\bfseries 01}
  (2016) 041},
\href{http://arxiv.org/abs/1511.03868}{{\ttfamily arXiv:1511.03868 [hep-th]}}.

\bibitem{D77}
M.~J. Duff, ``{Observations on Conformal Anomalies},''
\href{http://dx.doi.org/10.1016/0550-3213(77)90410-2}{{\em Nucl. Phys.}
  {\bfseries B125} (1977) 334}.

\bibitem{BCR83}
L.~Bonora, P.~Cotta-Ramusino, and C.~Reina, ``{Conformal Anomaly and
  Cohomology},''
\href{http://dx.doi.org/10.1016/0370-2693(83)90169-7}{{\em Phys. Lett.}
  {\bfseries B126} (1983) 305}.

\bibitem{BC77}
L.~S. Brown and J.~P. Cassidy, ``{Stress Tensor Trace Anomaly in a
  Gravitational Metric: General Theory, Maxwell Field},''
\href{http://dx.doi.org/10.1103/PhysRevD.15.2810}{{\em Phys. Rev.} {\bfseries
  D15} (1977) 2810}.

\bibitem{AGS05}
M.~Asorey, E.~V. Gorbar, and I.~L. Shapiro, ``{Universality and ambiguities of
  the conformal anomaly},''
  \href{http://dx.doi.org/10.1088/0264-9381/21/1/011}{{\em Class. Quant. Grav.}
  {\bfseries 21} (2003) 163--178},
\href{http://arxiv.org/abs/hep-th/0307187}{{\ttfamily arXiv:hep-th/0307187
  [hep-th]}}.

\bibitem{VFGS15}
A.~R. Vieira, J.~C.~C. Felipe, G.~Gazzola, and M.~Sampaio, ``{One-loop
  conformal anomaly in an implicit momentum space regularization framework},''
  \href{http://dx.doi.org/10.1140/epjc/s10052-015-3561-z}{{\em Eur. Phys. J.}
  {\bfseries C75} no.~7, (2015) 338},
\href{http://arxiv.org/abs/1505.05319}{{\ttfamily arXiv:1505.05319 [hep-th]}}.

\bibitem{Chu16}
C.-S. Chu and Y.~Koyama, ``{Adiabatic Regularization for Gauge Field and the
  Conformal Anomaly},''
\href{http://arxiv.org/abs/1610.00464}{{\ttfamily arXiv:1610.00464 [hep-th]}}.

\bibitem{Adler:1976zt}
S.~L. Adler, J.~C. Collins, and A.~Duncan, ``{Energy-Momentum-Tensor Trace
  Anomaly in Spin 1/2 Quantum Electrodynamics},''
\href{http://dx.doi.org/10.1103/PhysRevD.15.1712}{{\em Phys. Rev.} {\bfseries
  D15} (1977) 1712}.

\bibitem{Nielsen:1977sy}
N.~K. Nielsen, ``{The Energy Momentum Tensor in a Nonabelian Quark Gluon
  Theory},''
\href{http://dx.doi.org/10.1016/0550-3213(77)90040-2}{{\em Nucl. Phys.}
  {\bfseries B120} (1977) 212--220}.

\bibitem{MOM}
G.~Martinelli, C.~Pittori, C.~T. Sachrajda, M.~Testa, and A.~Vladikas, ``{A
  General method for nonperturbative renormalization of lattice operators},''
  \href{http://dx.doi.org/10.1016/0550-3213(95)00126-D}{{\em Nucl. Phys.}
  {\bfseries B445} (1995) 81--108},
\href{http://arxiv.org/abs/hep-lat/9411010}{{\ttfamily arXiv:hep-lat/9411010
  [hep-lat]}}.

\bibitem{Prince15}
J.~Gomis, P.-S. Hsin, Z.~Komargodski, A.~Schwimmer, N.~Seiberg, and S.~Theisen,
  ``{Anomalies, Conformal Manifolds, and Spheres},''
  \href{http://dx.doi.org/10.1007/JHEP03(2016)022}{{\em JHEP} {\bfseries 03}
  (2016) 022},
\href{http://arxiv.org/abs/1509.08511}{{\ttfamily arXiv:1509.08511 [hep-th]}}.

\bibitem{O91}
H.~Osborn, ``{Weyl consistency conditions and a local renormalization group
  equation for general renormalizable field theories},''
\href{http://dx.doi.org/10.1016/0550-3213(91)80030-P}{{\em Nucl. Phys.}
  {\bfseries B363} (1991) 486--526}.

\bibitem{FK09}
D.~Friedan and A.~Konechny, ``{Gradient formula for the beta-function of 2d
  quantum field theory},''
  \href{http://dx.doi.org/10.1088/1751-8113/43/21/215401}{{\em J. Phys.}
  {\bfseries A43} (2010) 215401},
\href{http://arxiv.org/abs/0910.3109}{{\ttfamily arXiv:0910.3109 [hep-th]}}.

\bibitem{N13}
Y.~Nakayama, ``{Scale invariance vs conformal invariance},''
  \href{http://dx.doi.org/10.1016/j.physrep.2014.12.003}{{\em Phys. Rept.}
  {\bfseries 569} (2015) 1--93},
\href{http://arxiv.org/abs/1302.0884}{{\ttfamily arXiv:1302.0884 [hep-th]}}.

\bibitem{F83}
M.~Freeman, ``{The Renormalization of Nonabelian Gauge Theories in Curved
  Space-time},''
\href{http://dx.doi.org/10.1016/0003-4916(84)90022-8}{{\em Annals Phys.}
  {\bfseries 153} (1984) 339}.

\bibitem{Stelle}
K.~S. Stelle, ``{Renormalization of Higher Derivative Quantum Gravity},''
\href{http://dx.doi.org/10.1103/PhysRevD.16.953}{{\em Phys. Rev.} {\bfseries
  D16} (1977) 953--969}.

\bibitem{DS93}
S.~Deser and A.~Schwimmer, ``{Geometric classification of conformal anomalies
  in arbitrary dimensions},''
  \href{http://dx.doi.org/10.1016/0370-2693(93)90934-A}{{\em Phys. Lett.}
  {\bfseries B309} (1993) 279--284},
\href{http://arxiv.org/abs/hep-th/9302047}{{\ttfamily arXiv:hep-th/9302047
  [hep-th]}}.

\bibitem{PZprep17}
V.Prochazka and R.Zwicky. In preparation, 2017.

\bibitem{Gukov:2015qea}
S.~Gukov, ``{Counting RG flows},''
  \href{http://dx.doi.org/10.1007/JHEP01(2016)020}{{\em JHEP} {\bfseries 01}
  (2016) 020},
\href{http://arxiv.org/abs/1503.01474}{{\ttfamily arXiv:1503.01474 [hep-th]}}.

\bibitem{LPR12}
M.~A. Luty, J.~Polchinski, and R.~Rattazzi, ``{The $a$-theorem and the
  Asymptotics of 4D Quantum Field Theory},''
  \href{http://dx.doi.org/10.1007/JHEP01(2013)152}{{\em JHEP} {\bfseries 01}
  (2013) 152},
\href{http://arxiv.org/abs/1204.5221}{{\ttfamily arXiv:1204.5221 [hep-th]}}.

\bibitem{LS89}
H.~Leutwyler and M.~A. Shifman, ``{GOLDSTONE BOSONS GENERATE PECULIAR CONFORMAL
  ANOMALIES},''
\href{http://dx.doi.org/10.1016/0370-2693(89)91730-9}{{\em Phys. Lett.}
  {\bfseries B221} (1989) 384--388}.

\bibitem{CCJ70}
C.~G. Callan, Jr., S.~R. Coleman, and R.~Jackiw, ``{A New improved energy -
  momentum tensor},''
\href{http://dx.doi.org/10.1016/0003-4916(70)90394-5}{{\em Annals Phys.}
  {\bfseries 59} (1970) 42--73}.

\bibitem{DV82}
M.~B. Voloshin and A.~D. Dolgov, ``{ON GRAVITATIONAL INTERACTION OF THE
  GOLDSTONE BOSONS},'' {\em Sov. J. Nucl. Phys.} {\bfseries 35} (1982)
  120--121.
[Yad. Fiz.35,213(1982)].

\bibitem{DL91}
J.~F. Donoghue and H.~Leutwyler, ``{Energy and momentum in chiral theories},''
\href{http://dx.doi.org/10.1007/BF01560453}{{\em Z. Phys.} {\bfseries C52}
  (1991) 343--351}.

\bibitem{K11}
Z.~Komargodski, ``{The Constraints of Conformal Symmetry on RG Flows},''
  \href{http://dx.doi.org/10.1007/JHEP07(2012)069}{{\em JHEP} {\bfseries 07}
  (2012) 069},
\href{http://arxiv.org/abs/1112.4538}{{\ttfamily arXiv:1112.4538 [hep-th]}}.

\bibitem{BKZR14}
F.~Baume, B.~Keren-Zur, R.~Rattazzi, and L.~Vitale, ``{The local
  Callan-Symanzik equation: structure and applications},''
  \href{http://dx.doi.org/10.1007/JHEP08(2014)152}{{\em JHEP} {\bfseries 08}
  (2014) 152},
\href{http://arxiv.org/abs/1401.5983}{{\ttfamily arXiv:1401.5983 [hep-th]}}.

\bibitem{CT13}
R.~J. Crewther and L.~C. Tunstall, ``{$\Delta I=1/2$ rule for kaon decays
  derived from QCD infrared fixed point},''
  \href{http://dx.doi.org/10.1103/PhysRevD.91.034016}{{\em Phys. Rev.}
  {\bfseries D91} no.~3, (2015) 034016},
\href{http://arxiv.org/abs/1312.3319}{{\ttfamily arXiv:1312.3319 [hep-ph]}}.

\bibitem{CT15}
R.~J. Crewther and L.~C. Tunstall, ``{Status of Chiral-Scale Perturbation
  Theory},'' {\em PoS} {\bfseries CD15} (2015) 132,
\href{http://arxiv.org/abs/1510.01322}{{\ttfamily arXiv:1510.01322 [hep-ph]}}.

\bibitem{Zoller:2016iam}
M.~F. Zoller, ``{On the renormalization of operator products: the scalar
  gluonic case},''
\href{http://arxiv.org/abs/1601.08094}{{\ttfamily arXiv:1601.08094 [hep-ph]}}.

\bibitem{C74}
W.~E. Caswell, ``{Asymptotic Behavior of Nonabelian Gauge Theories to Two Loop
  Order},''
\href{http://dx.doi.org/10.1103/PhysRevLett.33.244}{{\em Phys. Rev. Lett.}
  {\bfseries 33} (1974) 244}.

\bibitem{BZ81}
T.~Banks and A.~Zaks, ``{On the Phase Structure of Vector-Like Gauge Theories
  with Massless Fermions},''
\href{http://dx.doi.org/10.1016/0550-3213(82)90035-9}{{\em Nucl.Phys.}
  {\bfseries B196} (1982) 189}.

\bibitem{Zoller:2012qv}
M.~F. Zoller and K.~G. Chetyrkin, ``{OPE of the energy-momentum tensor
  correlator in massless QCD},''
  \href{http://dx.doi.org/10.1007/JHEP12(2012)119}{{\em JHEP} {\bfseries 12}
  (2012) 119},
\href{http://arxiv.org/abs/1209.1516}{{\ttfamily arXiv:1209.1516 [hep-ph]}}.

\bibitem{5loops}
P.~A. Baikov, K.~G. Chetyrkin, and J.~H. Kühn, ``{Five-Loop Running of the QCD
  coupling constant},''
\href{http://arxiv.org/abs/1606.08659}{{\ttfamily arXiv:1606.08659 [hep-ph]}}.

\bibitem{JO13}
I.~Jack and H.~Osborn, ``{Constraints on RG Flow for Four Dimensional Quantum
  Field Theories},''
  \href{http://dx.doi.org/10.1016/j.nuclphysb.2014.03.018}{{\em Nucl. Phys.}
  {\bfseries B883} (2014) 425--500},
\href{http://arxiv.org/abs/1312.0428}{{\ttfamily arXiv:1312.0428 [hep-th]}}.

\bibitem{Cardy:1988cwa}
J.~L. Cardy, ``{Is There a c Theorem in Four-Dimensions?},''
\href{http://dx.doi.org/10.1016/0370-2693(88)90054-8}{{\em Phys. Lett.}
  {\bfseries B215} (1988) 749--752}.

\bibitem{Christensen:1978md}
S.~M. Christensen and M.~J. Duff, ``{New Gravitational Index Theorems and
  Supertheorems},''
\href{http://dx.doi.org/10.1016/0550-3213(79)90516-9}{{\em Nucl. Phys.}
  {\bfseries B154} (1979) 301--342}.

\bibitem{Behr:2013vta}
N.~Behr and A.~Konechny, ``{Renormalization and redundancy in 2d quantum field
  theories},'' \href{http://dx.doi.org/10.1007/JHEP02(2014)001}{{\em JHEP}
  {\bfseries 02} (2014) 001},
\href{http://arxiv.org/abs/1310.4185}{{\ttfamily arXiv:1310.4185 [hep-th]}}.

\bibitem{BC80}
L.~S. Brown and J.~C. Collins, ``{Dimensional Renormalization of Scalar Field
  Theory in Curved Space-time},''
\href{http://dx.doi.org/10.1016/0003-4916(80)90232-8}{{\em Annals Phys.}
  {\bfseries 130} (1980) 215}.

\bibitem{GC1}
L.~Del~Debbio and R.~Zwicky, ``{Renormalisation group, trace anomaly and
  Feynman–Hellmann theorem},''
  \href{http://dx.doi.org/10.1016/j.physletb.2014.05.038}{{\em Phys. Lett.}
  {\bfseries B734} (2014) 107--110},
\href{http://arxiv.org/abs/1306.4274}{{\ttfamily arXiv:1306.4274 [hep-ph]}}.

\bibitem{GC2}
V.~Prochazka and R.~Zwicky, ``{Gluon condensates from the Hamiltonian
  formalism},'' \href{http://dx.doi.org/10.1088/1751-8113/47/39/395402}{{\em J.
  Phys.} {\bfseries A47} (2014) 395402},
\href{http://arxiv.org/abs/1312.5495}{{\ttfamily arXiv:1312.5495 [hep-ph]}}.

\bibitem{A01eta}
D.~Anselmi, ``{A Note on the improvement ambiguity of the stress tensor and the
  critical limits of correlation functions},''
  \href{http://dx.doi.org/10.1063/1.1475766}{{\em J. Math. Phys.} {\bfseries
  43} (2002) 2965--2977},
\href{http://arxiv.org/abs/hep-th/0110292}{{\ttfamily arXiv:hep-th/0110292
  [hep-th]}}.

\bibitem{LeeWick}
T.~D. Lee and G.~C. Wick, ``{Finite Theory of Quantum Electrodynamics},''
  \href{http://dx.doi.org/10.1103/PhysRevD.2.1033}{{\em Phys. Rev.} {\bfseries
  D2} (1970) 1033--1048}.
[,129(1970)].

\bibitem{GOW}
B.~Grinstein, D.~O'Connell, and M.~B. Wise, ``{The Lee-Wick standard model},''
  \href{http://dx.doi.org/10.1103/PhysRevD.77.025012}{{\em Phys. Rev.}
  {\bfseries D77} (2008) 025012},
\href{http://arxiv.org/abs/0704.1845}{{\ttfamily arXiv:0704.1845 [hep-ph]}}.

\bibitem{CZAKON}
M.~Czakon, ``{The Four-loop QCD beta-function and anomalous dimensions},''
  \href{http://dx.doi.org/10.1016/j.nuclphysb.2005.01.012}{{\em Nucl. Phys.}
  {\bfseries B710} (2005) 485--498},
\href{http://arxiv.org/abs/hep-ph/0411261}{{\ttfamily arXiv:hep-ph/0411261
  [hep-ph]}}.

\end{thebibliography}\endgroup

\end{document}